\documentclass[iop,revtex4]{emulateapj}
\pdfoutput=1
\usepackage{graphicx}
\usepackage{lscape}
\usepackage{url}
\newcommand{\logg} {\log g}
\newcommand{\Te} {T_{\rm eff}}
\newcommand{\mv} {$M_V$}
\newcommand{\msun} {$M_\odot$}
\newcommand\gta{\lower 0.5ex\hbox{$\buildrel > \over \sim\ $}} 
\newcommand\lta{\lower 0.5ex\hbox{$\buildrel < \over \sim\ $}} 

\newcommand{\ha} {$\rm{H}{\alpha}$}
\newcommand{\hb} {$\rm{H}{\beta}$}
\newcommand{\nuv} {${\rm{NUV}}-V$}

\begin{document}

\title{Toward a Spectroscopic Census of White Dwarfs Within \\40 parsecs of the Sun}

\author{M.-M. Limoges\footnote{Visiting Astronomer, Kitt Peak National Observatory, 
National Optical Astronomy Observatory, which is operated by the Association of 
Universities for Research in Astronomy (AURA) under cooperative agreement with the 
National Science Foundation.}, S. L\'epine$^2$, and P. Bergeron$^1$}
\affil{$^1$D\'epartement de Physique, Universit\'e de Montr\'eal, C.P.~6128, 
Succ.~Centre-Ville, Montr\'eal, Qu\'ebec H3C 3J7, Canada}

\affil{$^2$Department of Astrophysics, Division of Physical Sciences, 
American Museum of Natural History,
Central Park West at 79th Street, New York, NY 1002}
\email{limoges@astro.umontreal.ca, lepine@amnh.org, bergeron@astro.umontreal.ca}

\begin{abstract}
  We present the preliminary results of a survey aimed at
  significantly increasing the range and completeness of the local
  census of spectroscopically confirmed white dwarfs.  The current
  census of nearby white dwarfs is reasonably complete only to about
  20 parsecs of the Sun, a volume that includes around 130 white
  dwarfs, a sample too small for detailed statistical analyses. This
  census is largely based on follow-up investigations of stars with
  very large proper motions. We describe here the basis of a method
  that will lead to a catalog of white dwarfs within 40 parsecs of the
  Sun and north of the celestial equator, thus increasing by a factor
  of 8 the extent of the northern sky census. White dwarf candidates
  are identified from the SUPERBLINK proper motion database, allowing
  us to investigate stars down to a proper motion limit $\mu>40$ mas
  yr$^{-1}$, while minimizing the kinematic bias for nearby
  objects. The selection criteria and distance estimates are based on
  a combination of color-magnitude and reduced proper motion
  diagrams. Our follow-up spectroscopic observation campaign has so
  far uncovered 193 new white dwarfs, among which we identify 127 DA
  (including 9 DA+dM and 4 magnetic), 1 DB, 56 DC, 3 DQ, and 6 DZ
  stars.  We perform a spectroscopic analysis on a subsample of 84
  DAs, and provide their atmospheric parameters. In particular, we
  identify 11 new white dwarfs with spectroscopic distances within 25
  pc of the Sun, including 5 candidates to the $D<20$ pc subset.

\end{abstract}


\section{Introduction}

Statistics of the local white dwarf population, such as the space
density, luminosity function, and mass distribution, are fundamental
tools for understanding the evolution of the Galactic stellar
populations and quantifying their ages (\citealt{oswalt96},
\citealt{leggett98}). Because of their low luminosities, obtaining a
large and complete census of white dwarfs within a well-defined volume
remains a challenge. The best volume that can be defined for a census
of low-luminosity objects is the solar neighborhood, which alleviates
the need for deep surveys, and also allows one to map out the sample
in velocity space using readily available proper motions.

A catalog and analysis of the sample of white dwarfs within 20 pc of
the Sun were presented by \citet{holberg02}, and later refined by
\citet{holberg08} and \citet{sion09}. In light of these studies, the
current census of nearby white dwarfs is believed to be 80\% complete,
and contains 127 white dwarfs \citep{sion09}. Every white dwarf
suspected to lie within 20 parsecs of the Sun was analyzed in greater
detail by \citet{noemi2012}, and 130 members ended up in their sample
of local white dwarfs. Even if one assumes that the local sample is
complete, the size of the sample is too small for detailed statistical
analyses, and there is a need to extend the census and obtain a
complete sample of white dwarfs from a larger volume. Such an effort was
undertaken by \citet[][and earlier references within]{suba09} by
measuring trigonometric parallaxes for new white dwarfs that are
candidates of the 25 pc sample, as part of their DENSE project
focused on objects in the southern hemisphere. \citet{holberg2011}
also announced that the complete sample of white dwarfs will be
extended to 25 parsecs, thus doubling the volume of the local
sample. Based on the space density of white dwarfs known within 10 pc
of the Sun, \citet{suba09} estimated that the census of white dwarfs
within 25 pc {\em and} with accurate trigonometric parallaxes is only
$\sim40\%$ complete, and if we extend this horizon a little further
--- to 40 pc for instance --- the census of white dwarfs remains
largely incomplete.

Nearby white dwarfs have been traditionally discovered in catalogs of
stars with high proper motions. Major contributions have been made,
for instance, by \citet{LHS,NLTT}, \citet{gic1971}, and
\citet{giclas78}, who identified a significant number of faint, blue,
high proper motion stars, and their pioneer work is still useful to
today's astronomers. Indeed, in the first study dedicated to building
a complete census of the local sample of white dwarfs by
\citet{holberg02}, LHS, G, and GD objects form an important fraction
of the 109 objects reported in that sample. Major contributions to the
completeness of the local white dwarf sample also come from the work
of \citet{vennes2003}, \citet{kawka2004}, and \citet{kawka2006}, who
surveyed the revised NLTT catalog of \citet{salim2003}, and in
particular, identified eight new white dwarfs lying within 20 pc of
the Sun. The contribution of \citet{farihi2005} is also worth
mentioning in this effort, as well as those of
\citet{suba2007,suba2008,suba09}, and \citet{sayres2012}, aimed at
completing the 25 pc sample.

But in order to extend the volume of our complete sample of white
dwarfs, the first step is to identify nearby stars with smaller proper
motions, the coolest ($\Te\sim3500$ K) of which are extremely faint
due to the intrinsic small radius of white dwarfs. With the goal to
improve the statistics of the local white dwarf population, we have
been hunting for white dwarfs in the SUPERBLINK catalog.  This
catalog, which is based on a re-analysis of the Digitized Sky Surveys
--- with its 20-45 yr baseline --- is at least 95\% complete for the
entire northern sky down to $V=19.0$, with a very low rate of spurious
detection. It thus constitutes an ideal database from which to search
for faint, high proper motion objects such as nearby white
dwarfs. Also, because of its low proper motion limit ($\mu>0\farcs04$
yr$^{-1}$), the SUPERBLINK sample effectively eliminates the kinematic
bias for stars in the immediate vicinity of the Sun, which is a known
limitation of traditional catalogues such as the LHS catalog
($\mu>0\farcs5$ yr$^{-1}$; \citealt{LHS}) and the NLTT catalog
($\mu>0\farcs18$ yr$^{-1}$; \citealt{NLTT}).  Hence, the SUPERBLINK
catalog also represents a powerful tool for the study of the solar
neighborhood. Searching this database should provide a complete sample
of white dwarfs to a much larger distance limit.

Also, the high completeness and deep magnitude limit of SUPERBLINK
allows the detection of all white dwarfs down to the luminosity
function turnoff, which occurs at $L/L_{\odot}\simeq10^{-4}$
\citep{fontaine01}, up to a relatively large distance. For a 0.6 \msun\ white dwarf with a pure hydrogen
atmosphere, for instance, this corresponds to $\Te=5000$ K, or \mv=15.23.  The
limiting magnitude of $V=19$ implies that SUPERBLINK should be
detecting all white dwarfs down to the luminosity function turnoff to
a distance of 56.7 pc from the Sun.  The main question is what
fraction of these stars are expected to have proper motions above the
SUPERBLINK limit of $\mu>0\farcs04$ yr$^{-1}$. Assuming that the
distribution of velocities for white dwarfs to be the same as that of
main-sequence stars in the vicinity of the Sun, we can use Figure 1 of
\citet{LG2011}, which shows the kinematic selection effects of
SUPERBLINK by illustrating the fraction of stars in the
\emph{Hipparcos} catalog that would be selected with a proper motion
cut of $\mu>0\farcs04$ yr$^{-1}$ up to a given distance.  At 56.7 pc,
more than 90\% of the stars are detected. This minimal kinematic bias therefore 
allows one to detect most white dwarfs down to the luminosity function
turnoff, and to perform a complete statistical analysis on a sample
$\sim10$ times larger than the current 20 pc census. 

The interest in the local population of white dwarf stars in not only
statistical, but also astrophysical. Indeed, probing the solar
neighborhood allows the detection of faint, cool stars that
would remain undetected at larger distances. Since the cool end of the
white dwarf luminosity function is incomplete, obtaining a reliable
estimate of the space density of white dwarf stars and comparing the
luminosity function to models remains a challenge. The completion of
the cool end of the white dwarf luminosity function would allow the
accurate determination of the Galactic age and the verification of the
white dwarf cooling theory. Furthermore, many cool white dwarfs are
peculiar \citep{noemi2012}, and it is among them that we can expect to
find transition objects that would allow us to establish the link
between the different spectral types and to achieve a better
understanding of the white dwarf spectral evolution. The catalog of
\citet{holberg08} contains a large number of stars of particular
astrophysical interest. For a detailed description of these stars, see
\citet{noemi2012} and references therein. It is expected that surveys
at 25 and 40 parsecs will unveil an even larger number of such key
objects.

In this paper, we search the SUPERBLINK catalog to extend
significantly the census of white dwarfs in the solar neighborhood.
\citet{lspm05} have shown how reduced proper motion diagrams
constructed from the SUPERBLINK catalog can produce a large number
of white dwarf candidates. We present here a more detailed search and
identification of these white dwarfs through a large spectroscopic follow-up program. 
Our specific goal is to obtain 
spectral confirmation of all suspected white dwarfs within 40 parsecs
of the Sun. Given the enormous amount of data and limited telescope
access, we restrict ourselves to the northern part of the sky.  The
first step of this spectroscopic survey consists in the
identification, observation, and classification of the white dwarf
candidates. In Section 2, we present the catalog from which the
candidates are obtained. We detail our selection method in Section
3, as well as distance estimates and candidate list in Section
4. Section 5 describes the results of our spectroscopic observation
campaign, while a preliminary spectroscopic analysis, including the
determination of atmospheric parameters, is provided in Section
6. Finally, a discussion follows in Section 7. A more thorough model
atmosphere analysis of the atmospheric parameters of our complete
survey of new white dwarfs within 40 parsecs will be reported in subsequent papers.

\section{Proper Motion and Photometric Database}

Our white dwarf candidates are identified from the SUPERBLINK catalog
of stars with proper motions $\mu> 40$ mas yr$^{-1}$. This catalog,
based on a re-analysis of the Digitized Sky Surveys (which include POSS-I 
and POSS-II plate scans), is estimated to be $> 95\%$ complete in the
northern hemisphere down to a visual magnitude of $V=19$, 
but extends to $V\sim20$ in many areas of higher Galactic latitudes. 
The current version of the catalog comprises 2,283,540 objects, all
designated by the letters ``PM I'' followed by 10 characters based on
the right ascension ($\alpha$) and declination ($\delta$) of the
object. The basic search algorithm is described in \citet{lepine02},
while quality control procedures, including cross-correlation with
other catalogs and the compilation of astrometric and photometric
results, are discussed at length in \citet{lspm05} and \citet{LG2011}. 
A complete list of 61,977 northern stars with $\mu >
150$ mas yr$^{- 1}$ has already been published in \citet{lspm05}. We
provide below a brief summary of the astrometric and photometric
entries given in the current SUPERBLINK catalog.

\begin{deluxetable*}{lclcl}
\tablecolumns{5}
\tablewidth{0pt}
\tablecaption{Available Photometric Data}
\tablehead{
\colhead{Catalog} &
\colhead{Version} &
\colhead{Bands} &
\colhead{Counterparts} &
\colhead{Reference}}

\startdata

2MASS & $-$ &$JHK_S$  &1,472,665 & \citet{skrut06}\\
SDSS & DR6 & $ugriz$ & 345,958 & \citet{SDSS_DR6}\\
\emph{Hipparcos}, \emph{Tycho-2} & $-$ & $B_T, V_T$ & 118,000  & \citet{hipp2007}\\
& &  &  & \citet{hog} \\
USNO-B1.0 & $-$ & $B_J, R_F, I_N$ & 1,567,461& \citet{monet03}\\
GALEX & GR6 & FUV, NUV & 143,806 & \citet{gil09} \\

\enddata
\label{tab1}
\end{deluxetable*}

\subsection{Astrometry}

SUPERBLINK provides coordinates on the International Celestial
Reference System for the 2000.0 epoch. For stars catalogued in
\emph{Hipparcos}, the positions are extrapolated to the 2000.0 epoch
from the values given in \citet{hipp2007}, which are listed for the
1991.25 epoch.  Likewise, those not in \emph{Hipparcos} but listed in
Tycho-2 have their positions extrapolated from the proper motions
listed in Tycho-2, and if a star has a counterpart in 2MASS
\citep{Cutri2003}, its position is extrapolated from the position of
the 2MASS counterpart. Finally, coordinates for stars without a
\emph{Hipparcos} or 2MASS counterparts are calculated by SUPERBLINK
from the position of the stars on the POSS-II scans. The coordinates
of those stars are thus less accurate but are generally within a few 
arcseconds (see \citealt{lspm05} for details).

SUPERBLINK also lists proper motions for each entry, tabulated from
three sources.  When available, proper motions are taken from the
\emph{Hipparcos} catalog \citep{hipp2007} or from the \emph{Tycho-2}
catalog \citep{hog}.  Otherwise, the proper motions listed are those
measured in the SUPERBLINK proper motion survey, based on the
Digitized Sky Survey images. SUPERBLINK ends up providing proper
motions for more than 2 million stellar objects, and in particular, a
total of 1,567,461 stars with
$\delta>0$.  From now on, when we mention the SUPERBLINK database, we
refer to the northern part of the catalog.

\subsection{Photometric Data}

The construction of reduced proper motion diagrams requires, in
addition to proper motion measurements, a set of photometric data in
order to estimate the color of each star. Fortunately, the
cross-correlation of SUPERBLINK with other catalogs not only allows
coordinates and proper motions to be measured with more accuracy, but
it also provides a useful set of photometric data covering a large
portion of the electromagnetic spectrum. We describe these data in
turn, and a summary is provided in Table \ref{tab1}.

The Two Micron All Sky Survey (2MASS) Point Source Catalog
\citep{skrut06} represents an excellent source of near-infrared
magnitudes for our targets in SUPERBLINK since the 2MASS survey covers
the whole sky and is complete down to $J \sim 16.5$. \citet[][see
their Figure 30]{lspm05} successfully showed that white dwarfs in
SUPERBLINK could easily be separated from other stellar populations in
a $H_V$ vs $V-J$ reduced proper motion diagram.  For the present
study, we used a version of the SUPERBLINK catalog in which 2MASS
counterparts had already been found and assigned to 1,472,665 of the
stars ($\sim94\%$), with the remainder having no detectable
counterpart in 2MASS. These infrared $J$, $H$, and $K_S$ magnitudes
have a 0.02-0.03 mag accuracy down to 13th magnitude, and point
sources are detected with S/N better than 10 for stars brighter than
$J=15.9$, $H=15.0$, and $K_S=14.3$ \citep{skrut06}.

The Sloan Digital Sky Survey (SDSS) also represents a useful source of
photometric data, with $ugriz$ photometry from the Data Release 6
\citep{SDSS_DR6} for 345,958 counterparts in the SUPERBLINK catalog. The SDSS magnitudes have
photometric uncertainties of roughly 1\% in the $griz$ bands and 2\%
in $u$ \citep{Pad2008}. This is by far the most accurate optical
photometry available in our study, and will be especially useful to
identify white dwarfs in the SUPERBLINK catalog.

Optical photometry in the blue ($B_T$) and in the visual ($V_T$) range
are also extracted for 118,475 stars with counterparts in the
\emph{Hipparcos} and \emph{Tycho-2} catalogs. Additional optical
photometry was also obtained from the USNO-B1.0 database
\citep{monet03}, providing photographic magnitudes for the totality of
the catalog, i.e.~1,567,461 objects. However, for some entries, the
photometry is available only for one or two bands. More specifically,
$B_J$ magnitudes are available for 1,390,471 objects, $R_F$ magnitudes
for 1,405,840, and $I_N$ magnitudes for 912,550 objects.  The blue
$B_J$ magnitudes are extracted mostly from scans of IIIaJ plates from
the Palomar Sky Surveys (POSS-I, POSS-II) and the Southern ESO Schmidt
(SERC) Survey, the red $R_F$ magnitudes are extracted from scans of
IIIaF plates from POSS-I and POSS-II and also from the
Anglo-Australian Observatory red survey (AAO-red), while the
near-infrared $I_N$ magnitudes are extracted from the IVn plates from
POSS-II and SERC. The $B_T$ and $V_T$ magnitudes are more accurate
(0.1 mag or better) than the photographic magnitudes (typically 0.5
mag), but they are available only for the brightest stars in SUPERBLINK, 
while photographic magnitudes are available for every object.

Finally, we also searched the sixth data release (GR6) of the GALEX
database \citep{gil09} and identified 147,096 counterparts to the
SUPERBLINK objects (for a $5\farcs0$ search radius). The corresponding
far-ultraviolet (FUV, 1350-1780 \AA) and near-ultraviolet (NUV,
1770-2730 \AA) magnitudes are particularly useful for the
identification of blue objects, and in particular white dwarf stars.

\section{Selection of the Candidates Based on Reduced Proper Motion Diagrams} 

Reduced proper motion diagrams (RPMD) are a particularly efficient tool to 
identify white dwarf candidates with known proper motions (see, for
instance, \citealt{knox1999}, \citealt{oppen2001},
\citealt{vennes2003}, \citealt{carollo2006}, \citealt{kilic06}). The
reduced proper motion of an object is defined as $H_m=m+5\log\mu+5$,
where $m$ is the apparent magnitude in some bandpass and $\mu$ is the
proper motion measured in arcseconds per year. The reduced proper motion
is analogous to the absolute magnitude $M_m=m+5\log\pi+5$, where the
trigonometric parallax $\pi$ is replaced with the proper motion $\mu$
of the object.  A reduced proper motion diagram is thus similar to a
color-magnitude diagram, and white dwarfs occupy a similar location in
the diagram, i.e.~the bottom-left region.  Furthermore, using the
tangential velocity $v_{\rm tan}=4.74\,\mu \pi^{-1}$ in units of
$\rm{km}~s^{-1}$ instead of the proper motion, we obtain
$H_m=M_m+5\log{v_{\rm tan}}-3.38$, and each star population can be
isolated based on the mean value of its tangential velocity. 

One major problem with the identification of white dwarf candidates
using reduced proper motion diagrams is the contamination of the white
dwarf region by other stellar populations, and by high-velocity
subdwarfs in particular. \citet{vennes2003} showed, however, that this
contamination can be substantially reduced by the inclusion of a
criterion based on $V-J$. Similarly, \citet{kilic06} demonstrated that
reduced proper motion diagrams are efficient for detecting cool white
dwarfs only when the measured proper motions of all stellar
populations are reliable, since subdwarfs with inaccurate proper
motions can contaminate the other stellar populations, and notably the
white dwarf region of the diagram.  SUPERBLINK has an estimated false
detection level of less than $1\%$ down to $V=19$, but the false
detection rate increases significantly for fainter sources.  In our
selection criteria, we thus restrict our search to that stars with
$V<19$.  Fortunately, SUPERBLINK has a very high level of completeness
for $V<19$, exceeding $98\%$ for most of the sky. We are thus
confident that we can easily identify a significant fraction of the nearby
white dwarfs using this technique. The next sections describe the four
reduced proper motion diagrams we used to identify white dwarf candidates
in SUPERBLINK, in an effort to take advantage of the whole set of
photometric information available.  The order in which these are presented 
follows the order of their estimated efficiency at isolating the white dwarf 
population, starting with the most efficient one.

\subsection{RPMD Using $ugriz$ Photometry}

\begin{figure}[t!]
\epsscale{1.4}
\plotone{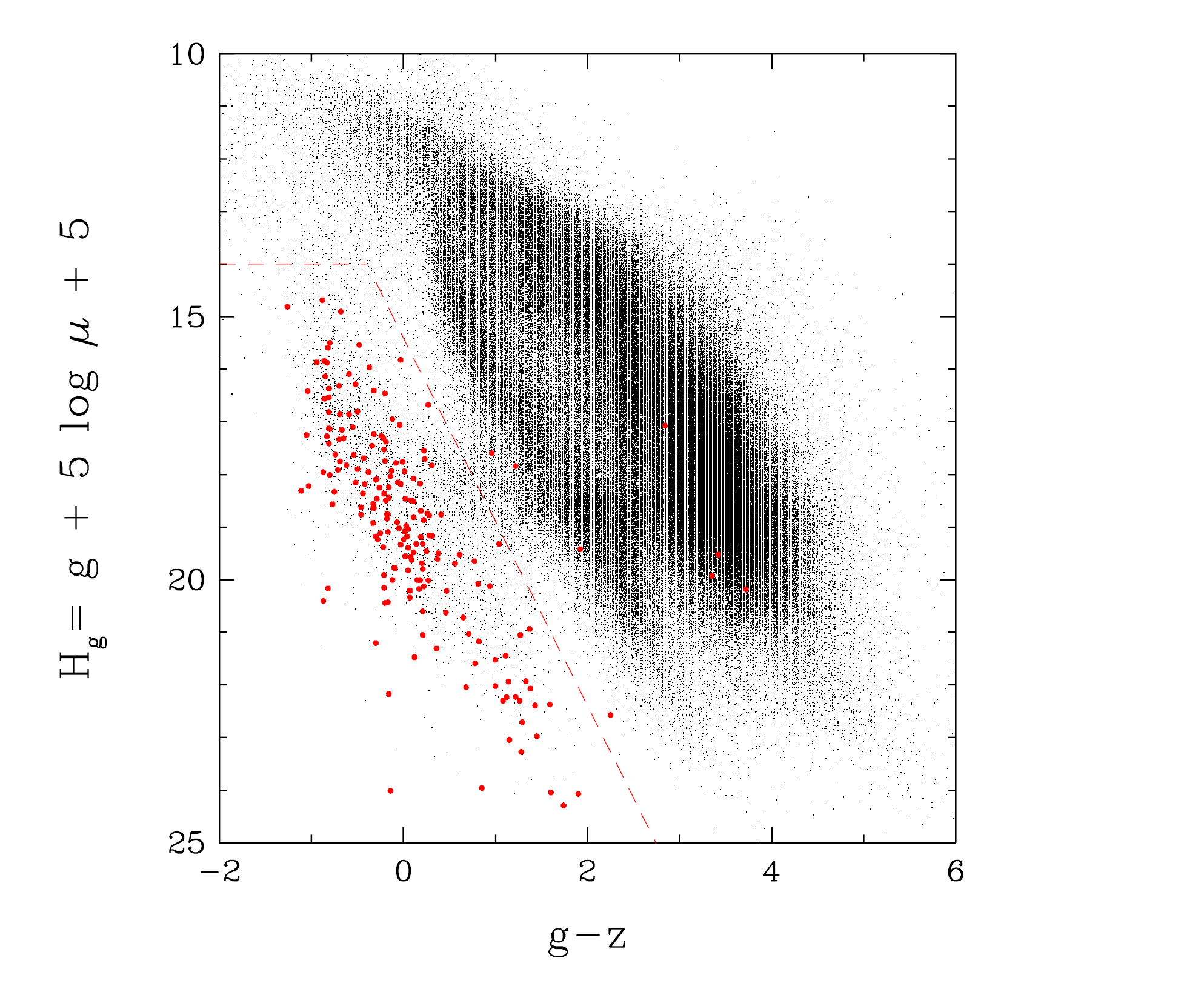}
\caption{Reduced proper motions diagram ($H_g, g-z$)
  for the 345,958 stars in the SUPERBLINK catalog ($\mu>40$ mas
  yr$^{-1}$) with counterparts in the SDSS 6th Data Release. The red
  dashed lines show the limits used to define the location of the
  white dwarfs. These limits include 4929 white dwarf candidates,
  i.e.~$1.4\%$ of the sample of SUPERBLINK stars with SDSS
  photometry. The spectroscopically confirmed white dwarfs from the WD
  Catalog are shown as large red dots. \label{fig1}}
\end{figure}

We present in Figure \ref{fig1} the $H_g$ vs $g-z$ reduced proper motion diagram
constructed from the 345,958 SUPERBLINK objects in the northern
hemisphere with $ugriz$ photometry available from the 6th Data Release
of the Sloan Digital Sky Survey. As a result of the relatively high
accuracy of the SDSS magnitudes, the white dwarf population is
particularly well separated from the other populations in this
diagram.

To verify the accuracy of our procedure, we also display as red dots
in Figure \ref{fig1} the sample of white dwarfs taken from the
2008 May electronic version of the Catalogue of Spectroscopically
Identified White
Dwarfs\footnote{http://www.astronomy.villanova.edu/WDCatalog/index.html}
\citep[][hereafter WD Catalog]{mccook99} with $\delta>0$ also found in
SUPERBLINK and with $g$ and $z$ photometry available. We first note
that 191 spectroscopically confirmed white dwarfs lie in the expected
region near the bottom left of the diagram, and that a small number of
white dwarfs are color outliers. More precisely, 3 white dwarfs
overlap the subdwarf region, and 4 are found in the redder, main
sequence portion of the diagram. Two of these are actually binary systems:
0855+604.1 is a DBQ \citep{green69} and 0855+604.2 is a DCE?
\citep{eggen1965}, while 1133+358 is an unresolved DC+dM
\citep{greenstein1976}.

Fortunately, there are enough spectroscopically identified white
dwarfs that are well separated from the other populations to allow us
to define selection criteria for the white dwarf area. These criteria
are defined by the need to include as many white dwarf candidates as
possible, while trying to keep the contamination from subdwarfs to a
minimum. As a general criterion, the area occupied by the white dwarfs
must include {\it at least} 80\% of the spectroscopically confirmed
white dwarfs. In the present case, this limit between the halo
subdwarfs and the white dwarf area is defined by the following linear
equation $H_g=3.5(g-z)+16.5$. The slope and y-intercept are chosen in
a very conservative manner, and include 96\% of the WD Catalog sample
with measured $ugriz$ photometry. The reason why we recover almost all
white dwarfs from the WD Catalog is that the known white dwarf
population is well separated in this reduced proper motion diagram,
and only a few objects are color outliers. However, in the upper part
of the diagram ($H_g\lesssim14$) there is again some confusion between
the different stellar populations defined by our linear equation, so
we simply apply an additional cutoff at $H_g=14$ based on the known
white dwarf population, in order to keep the contamination to a
minimum.  These adopted selection criteria for the $H_g$ vs $g-z$
diagram are displayed in Figure \ref{fig1}. Out of the 345,958
SUPERBLINK stars with a counterpart in the 6th Data Release of the
SDSS, about 4929 fall within the white dwarf selection limits. 
This number represents 1.4\% of
the stars in the catalog with SDSS photometry, or $0.5\%$ of
the total number of objects in SUPERBLINK.

\subsection{RPMD Using GALEX Photometry}
\begin{figure}[h!]
\epsscale{1.4}
\plotone{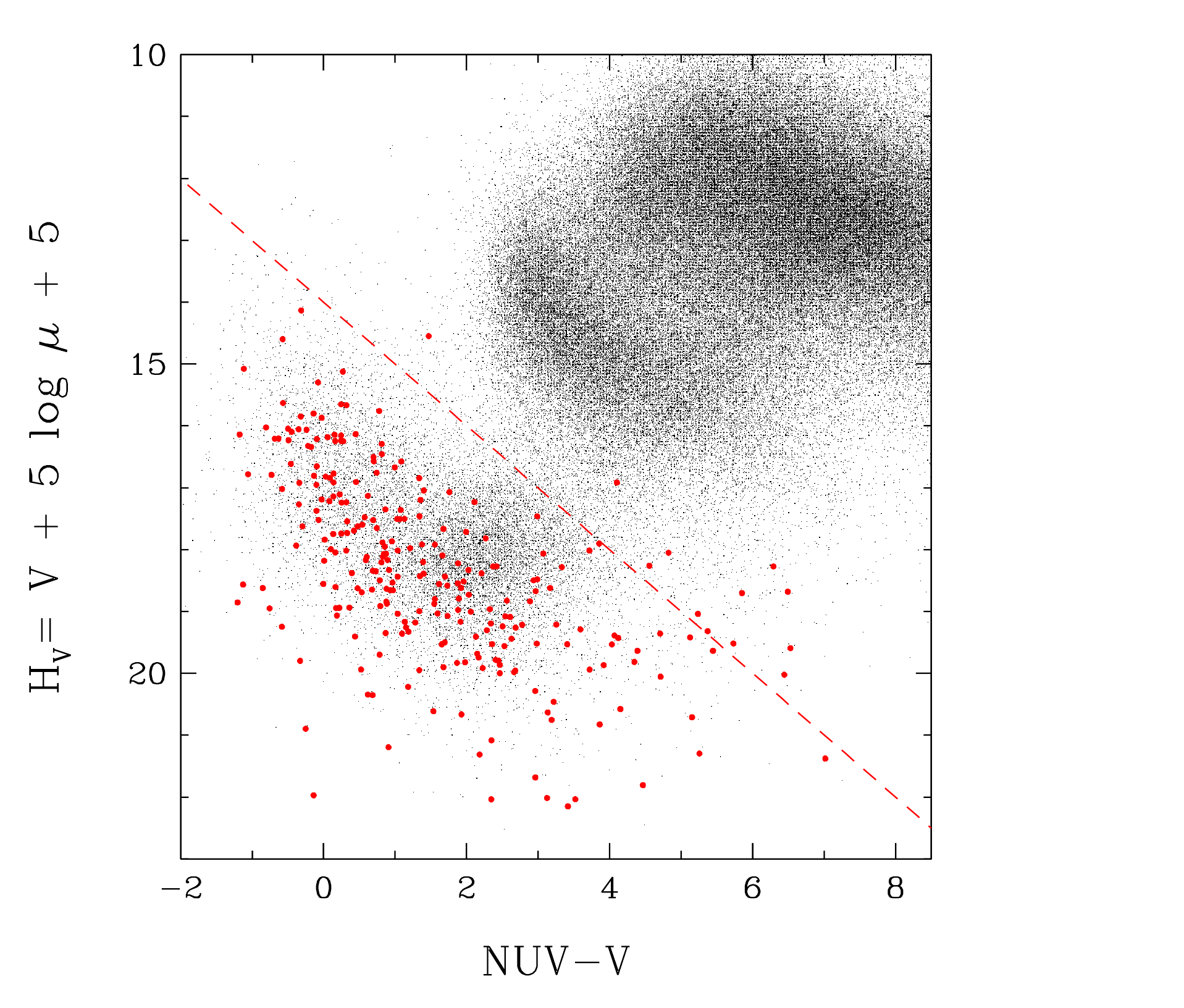}
\begin{flushright}
\caption{Reduced proper motions diagram ($H_V$, \nuv)
  for the 147,096 stars in the SUPERBLINK catalog ($\mu>40$ mas
  yr$^{-1}$) with counterparts in the 6th Data Release of the GALEX
  survey. The red dashed line shows the limits used to define the
  location of the white dwarfs. This limit includes 19,150 white dwarf
  candidates, i.e.~$12.7\%$ of the sample of SUPERBLINK stars with
  GALEX photometry. The spectroscopically confirmed white dwarfs from
  the WD Catalog are shown as large red dots. \label{fig2}}
\end{flushright}
\end{figure}
White dwarf stars are generally hotter than main sequence or subdwarf
stars, and since their atmospheres are usually devoid of heavy
elements that could absorb the UV flux, they are moderately strong UV
emitters and can easily be distinguished from non-degenerate stars in
a reduced proper motion diagram built from GALEX photometry. We
present in Figure \ref{fig2} the $H_V$ vs \nuv\ diagram containing
147,096 stars in SUPERBLINK with NUV magnitudes measured by GALEX. NUV
magnitudes are used in this diagram since they are available for a
much larger number of stars than the FUV magnitudes.  No corrections
are applied for interstellar reddening, since according to the
characterization of the Local Bubble of \citet{Reis2011}, the smallest
distance to the wall of dust that causes extinction ($E(b-y)\ge0.040$)
is $\sim80$ pc. The reddening should thus not affect the white dwarf
candidates of the local sample.

Here and in the following diagrams, the $V$ magnitudes are estimated
from the relation

\begin{equation}
V=B_J-0.46(B_J-R_F) 
\end{equation} 

\noindent
as recommended by \citet{lspm05}, where $B_J$ and $R_F$ are
photographic magnitudes taken from the USNO-B1.0 catalog. These
estimated $V$ magnitudes are believed to be accurate to $\pm 0.5$ mag
\citep{lspm05}. As mentioned earlier, while $V_T$ magnitudes from the
\emph{Tycho-2} catalog are more accurate than photographic magnitudes,
they are only available for a small number of SUPERBLINK objects,
whereas $B_J$ and $R_F$ exist for the majority of our targets.  Hence,
despite the relatively large uncertainties of the $V$ magnitudes
employed here, they have the advantage of being available for most of
the entries in SUPERBLINK.

As in Figure \ref{fig1}, the spectroscopically confirmed white dwarfs
from the WD Catalog are also shown in red. Here, a total of 13
spectroscopically confirmed white dwarfs white dwarfs are scattered in
areas normally occupied by other stellar populations. Most of these
objects are cool degenerates, including 9 DA, 2 DC, 1 DZ (1705+030,
Greenstein 1984) and 1 DQ star (1105+412, Koester and Knist, 2006).
Most likely, these have very little UV flux and thus inaccurate NUV
magnitudes.  

We define the slope and y-intercept of the line that characterizes the
white dwarf region with the same criteria as before. Also, since the
halo subdwarfs and main sequence stars are generally not as bright in
the UV as white dwarfs are, there is no need to apply any further cutoff
in $H_V$.  Therefore, to be considered a white dwarf candidate, a star
must have a reduced proper motion larger than $H_V=({\rm
  NUV}-V)+14$. With this limit, $94\%$ of the white dwarfs from the WD
Catalog with measured NUV photometry are recovered.  Finally, if this
criterion is applied to the 147,096 SUPERBLINK objects with both NUV
and $V$ magnitudes, we obtain a sample of 19,150 white dwarf
candidates, which represent 12.7\% of the stars with GALEX 
photometry in SUPERBLINK, or $2.0\%$ of the complete catalog.

\subsection{RPMD Using 2MASS Photometry}
\begin{figure}[h!]
\epsscale{1.4}
\plotone{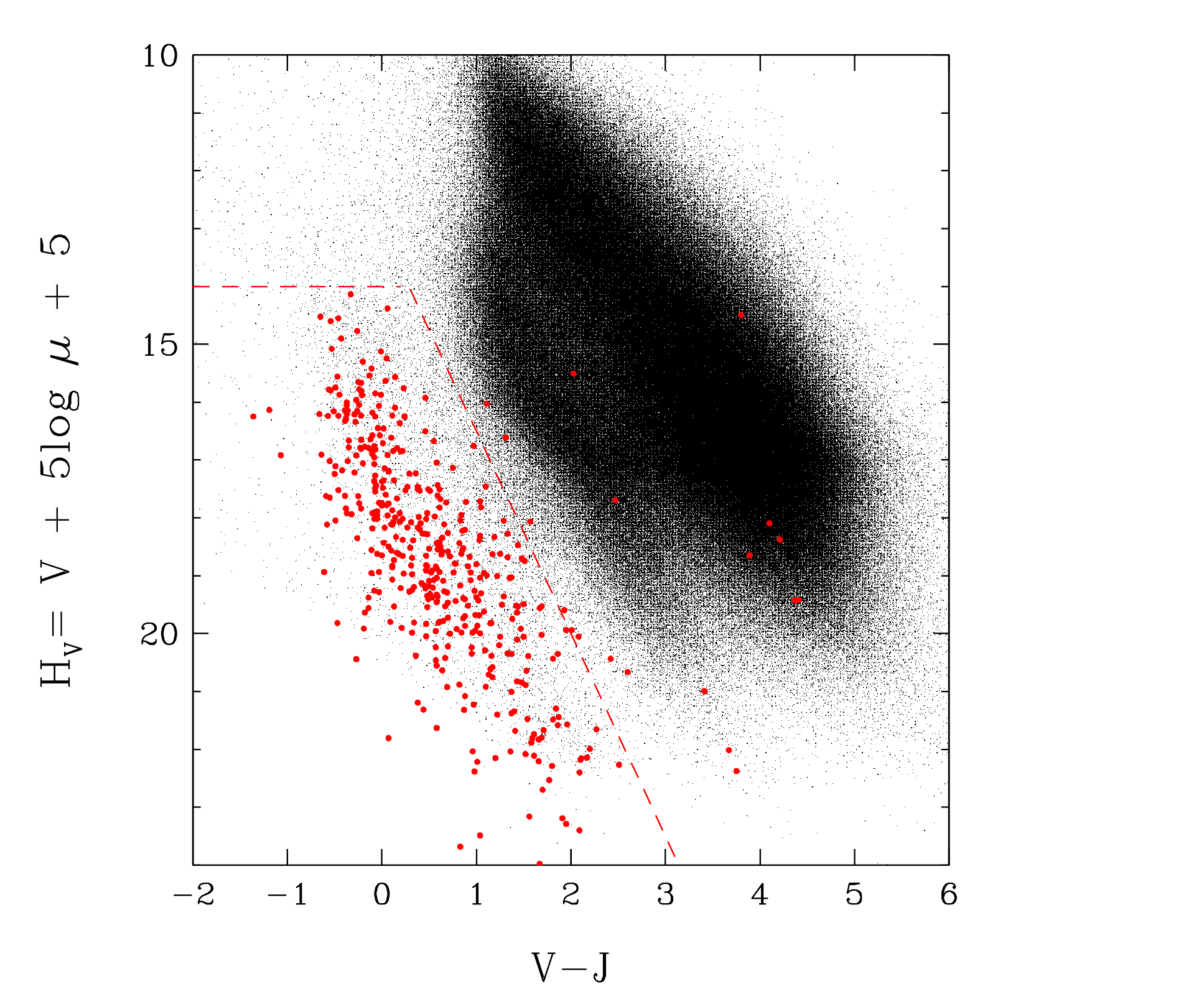}
\caption{Reduced proper motions diagram ($H_V, V-J$)
  for the 1,265,733 stars in the SUPERBLINK catalog ($\mu>40$ mas
  yr$^{-1}$) with counterparts in the 2MASS catalog and for which an
  estimation of the $V$ magnitude was possible. The red dashed lines
  show the limits used to define the location of the white
  dwarfs. These limits include 16,977 white dwarf candidates,
  i.e.~$1.3\%$ of the initial sample. The spectroscopically confirmed
  white dwarfs from the WD Catalog are shown as large red
  dots. \label{fig3}}
\end{figure}

As discussed above, the reduced proper motion diagram in $H_V$ vs
$V-J$, displayed in Figure 30 of \citet{lspm05}, showed that nearly
2000 white dwarfs could be identified in the SUPERBLINK catalog of
stars with proper motions $\mu>150$ mas yr$^{-1}$. Since the
SUPERBLINK catalog now includes stars with $\mu>40$ mas yr$^{-1}$, and
because a significant fraction of its entries has a counterpart in the
2MASS catalog, such a diagram has an even greater potential for
identifying white dwarf stars. The resulting $H_V$ vs $V-J$ diagram is
shown in Figure \ref{fig3}, and contains 1,265,733 stars from
SUPERBLINK with $J$ magnitudes taken from 2MASS as well as $B_J$ and
$R_F$ magnitudes from the USNO-B1.0 catalog. Here again, the $V$
magnitudes are obtained from the empirical calibration given by
Equation (1).

The comparison with the spectroscopically confirmed white dwarfs from
the WD Catalog reveals 20 objects that fall in a region to the right
of that occupied by the bulk of white dwarfs.  Among them, there
are 9 DA, 4 DC, 1 DQ, and 1 DZ star.  This diagram also includes the
largest number of multiple-star systems identified so far. Indeed, we
find four WD+dM systems, a DB+DC binary (2058+342,
\citealt{farihi04}), and 0023+388, the triple star system discussed
earlier.

Applying similar criteria as before, the white dwarf candidates are
selected from Figure \ref{fig3} if their reduced proper motion is
larger than $H_V=3.5(V-J)+13$, with a cutoff of $H_V>14$. These
limits, displayed in Figure \ref{fig3}, recover 86\% of the white
dwarfs from the WD Catalog listed in SUPERBLINK and for which 2MASS
and photographic magnitudes are available. This defines a sample of 
16,977 white dwarf candidates out of the 1,265,733 objects in the initial
SUPERBLINK sample with 2MASS photometry, or a fraction of $1.3\%$ 
(0.23\% of the total initial sample). 

\subsection{RPMD Using Photographic Magnitudes}
\begin{figure}[h!]
\epsscale{1.4}
\plotone{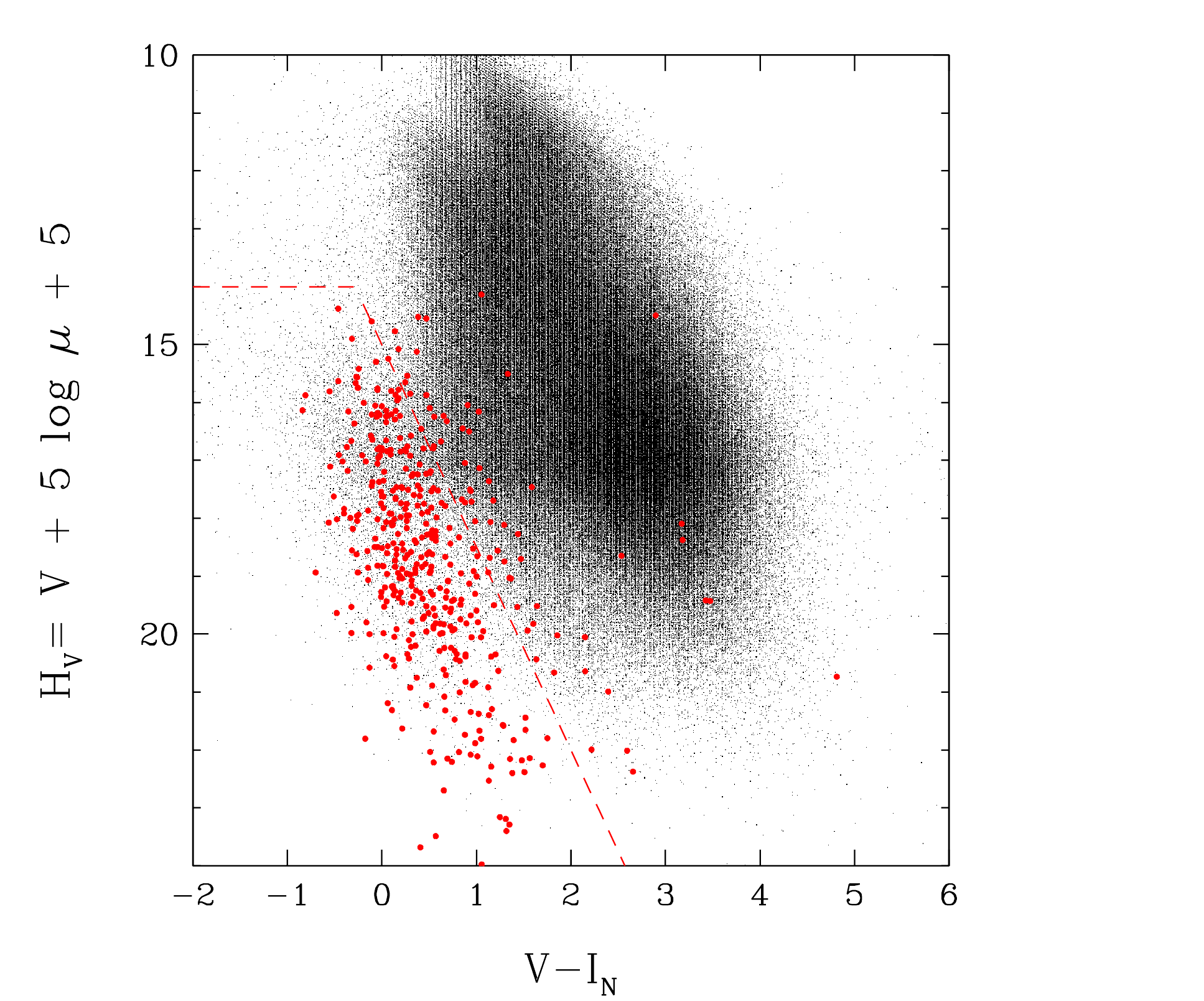}
\caption{Reduced proper motions diagram ($H_V,
  V-I_N$) for the 878,847 stars in the SUPERBLINK catalog ($\mu>40$
  mas yr$^{-1}$) with counterparts in the USNO-B1.0 catalog for all
  three $B_J$, $R_F$ and $I_N$ photographic magnitudes. The red dashed
  lines show the limits used to define the location of the white
  dwarfs. These limits include 20,862 white dwarf candidates,
  i.e.~$2.4\%$ of the initial sample. The spectroscopically confirmed
  white dwarfs from the WD Catalog are shown as large red
  dots. \label{fig4}}
\end{figure}

Even if USNO-B1.0 photographic magnitudes are less accurate than CCD
photometry obtained in recent large surveys such as SDSS, they have
the advantage of covering the whole sky, and are thus available for
all 1.6 million SUPERBLINK objects. We present in Figure
\ref{fig4} the reduced proper motion diagram in $H_V$ vs $V-I_N$
constructed with $V$ as defined by Equation (1). To be included in
this diagram, photographic $B_J$, $R_F$, and $I_N$ magnitudes must
all be available for each object. However, as discussed by
\citet{lspm05}, not all USNO-B1.0 entries have magnitude information
in all three bands. Whenever possible, L\'epine \& Shara tried to
combine data if a USNO-B1.0 star appeared as more than one entry, but
some sources remained without information for one or more bands. As a
result, Figure \ref{fig4} includes $878,847$ stars
out of the $1,567,461$ entries in SUPERBLINK with at least one 
photographic magnitude in USNO-B1.0.

In Figure \ref{fig4}, the separation between the white dwarf and
subdwarf populations is not as well defined as in the other reduced
proper motion diagrams.  The $V$ and $I_N$ filters are indeed too
close in wavelength to allow an efficient separation of the two
populations, as was the case with $V-J$ or $g-z$, for instance. We
also have to consider the fact that there is an uncertainty of
$\pm0.5$ mag in both $V$ and $I_N$, which adds to the dispersion in $V-I_N$. The
comparison of the white dwarfs in the WD Catalog displayed in Figure
\ref{fig4} with those in Figure \ref{fig3} using $V-J$
reveals that most of the 55 outliers in $V-I_N$ are also outliers in
$V-J$, the difference between the two diagrams being that there is a
larger concentration of outliers in $V-I_N$ near the white dwarf
locus. Once again, the measurement uncertainty in $V-I_N$ is to
blame. All of the interesting outliers have been discussed in the
preceding sections and will not be repeated here.

Despite this large contamination, we must still define some selection
criteria, and as before, the slope and y-intercept of the limit
between white dwarfs and halo subdwarfs is chosen to include $80\%$ of
the white dwarfs from the WD Catalog; a lower limit in $H_V$ is also
defined. Using these criteria, the white dwarf candidates are selected
from Figure \ref{fig4} if their reduced proper motion is larger
than $H_V=3.5(V-I_N)+13$, with a cutoff of $H_V>14$.  This time, since
the contamination is much larger than in previous reduced proper
motion diagrams, a more conservative cut must be used, and no more
than $80\%$ of the white dwarfs from the WD Catalog are recovered.  We
finally end up with 20,862 white dwarf candidates out of the $878,847$
SUPERBLINK objects in the original sample with USNO photometry, or a
fraction of $2.2\%$ of the total catalog.

Combining the results from all four reduced proper motion diagrams, we
obtain a total of 20,862 white dwarf candidates, since all of them
have USNO photographic magnitudes, but only a fraction of them have
data available in other photometric systems. Some of these candidates,
however, can be found in up to four reduced proper motion diagrams. At
this point, each candidate can be assigned an order of priority
depending on the quality of the photometry used for its
identification.

\subsection{A Priority Approach}

For each given star, data can be available for up to four photometric
systems and their corresponding reduced proper motion diagrams.  Thus,
it is possible that a star could be within the white dwarf region
defined by our selection criteria in one diagram and outside in
another diagram.  We must therefore decide which photometric system
should be prioritized to decide whether or not a star should be included 
in our final list of $\sim
21,000$ white dwarf candidates.  For instance, $ugriz$ magnitudes
should take precedence over photographic magnitudes, since the former
are much more accurate. In the following, we establish the order of
priority for our four photometric systems, based on their degree of
photometric accuracy.

SDSS magnitudes have a relatively high degree of accuracy and cover a
wavelength range that allows an efficient separation of the white
dwarfs from the other stellar populations, as discussed above.  The
reduced proper motion diagram based on $ugriz$ magnitudes is arguably
the most accurate, and it is therefore given the highest priority.
However, a comparison of our preliminary list of white dwarf
candidates based on $ugriz$ data with those found in the literature
shows a significant contamination from subdwarfs. \citet{kawka2004}
successfully reduce this contamination by including a color-color
criterion to their reduced proper motion cut, namely
$(V-J)<3.28(J-H)-0.75$. But such limits also get rid of the white
dwarfs with red, main-sequence companions, which we would like to
include.  Also, as observed by \citet{kilic06} and \citet{sayres2012},
any uncertainty in the proper motion measurements can lead to
contamination.  Also, spurious proper motions are possible for stars
near the faint magnitude limit of the SUPERBLINK.  With all these
considerations in mind, we apply on top of the criteria in ($H_g,
g-z$), a criterion inspired by \citet{kawka2004}, but based on Figure
\ref{fig3}. It is defined as follows: stars with $V<14$ must also have
$H_V=3.5(V-J)+13$. This limit on $V$ removes the largest number of
known main sequence stars from our list of candidates, while keeping
all the known white dwarfs. Applying the criterion in $(H_V, V-J)$ to
fainter candidates results in the elimination of spectroscopically
confirmed white dwarfs, so it is not applied to those fainter
stars. Finally, using the criteria defined in Section 3.1, combined
with those defined in Section 3.3 when $V<14$, we retain a total of
4823 white dwarf candidates. Unfortunately, the SDSS survey, at the
epoch of the DR6, only covered about a quarter of the northern
sky. Bright stars also tend to be saturated in SDSS, and we need to
limit our selection to stars with $u>13$, $g>14$, $r>14$, $i>14$, and
$z>12$ \citep{york}. Consequently, there is no SDSS counterpart for
every star in the SUPERBLINK catalog, and we must therefore rely on
other photometric systems to select our targets.

In the absence of SDSS photometric data, GALEX UV photometry is used
instead for the selection, whenever it is available. It is our second
choice because of the corresponding photometric accuracy as well as
the efficiency of the criteria in \nuv\ to separate white dwarfs from
other stellar populations. Here, we also apply the ($H_V, V-J$)
criterion for stars with $V<14.0$ in an effort to decontaminate our
sample of candidates. The criteria and the method described in Section
3.2 led to the identification of 8092 additional white dwarf
candidates. The GALEX survey covers $80\%$ of the sky, with special
care taken to avoid the Galactic plane and Magellanic clouds, which
could provide excess background flux in the
UV\footnote{http://galex.stsci.edu/GR6/}.

In the absence of SDSS and GALEX data, 2MASS photometry combined with
the criteria defined in Section 3.3 are used to identify 1132
additional white dwarf candidates with $\delta>0$ and $V<14.0$. The
2MASS survey is a precious source of photometric information since it
covers practically the whole sky. However, since it is only complete
down to $J\sim16.5$, our fainter targets do not have a 2MASS
counterpart.  Finally, if there is no SDSS, GALEX, or 2MASS photometry
available for a given target, we must rely on photographic magnitudes
obtained from the USNO-B1.0 catalog. With the criteria defined in
Section 3.4 and for $V<14.0$, we identify an additional list of 6688
white dwarf candidates.

All in all, a total of 20,735 white dwarf candidates are identified
with the help of the four reduced proper motion diagrams described in
this section. This large number of candidates amply justifies our
decision to restrict our search to the northern hemisphere. Moreover,
given our interest in establishing a census of white dwarfs in the
solar neighborhood within 40 pc of the Sun, we must further restrain
our list of candidates by evaluating photometric distances for each
object on our target list.

\section{A Sample of White Dwarf Candidates within 40 pc of the Sun}

\subsection{Distances from Color-Magnitude Relations} 

In the absence of trigonometric parallax measurements for most white
dwarf candidates in our sample, we must rely on distances estimated
using the only information available, which are the apparent
magnitudes of each star in a set of bandpasses covering the
ultraviolet to the near-infrared. In this section, we estimate {\it
  photometric distances} from the distance modulus, $m-M_m=5\log D-5$,
where the absolute magnitude $M_m$ is determined from theoretical
color-magnitude relations combined with a measured color index in some
specified photometric system. To do so, we rely on synthetic colors
obtained using the procedure outlined in \citet{HB06} based on the
improved Vega fluxes taken from \citet{bohlin2004}. These
color-magnitude relations are available on our Web site\footnote{See
  http://www.astro.umontreal.ca/\~{}bergeron/CoolingModels}, or upon
request on any photometric system. They cover a range of effective
temperature between $\Te = 1500$~K and 110,000~K and surface gravities
between $\logg=6.5$ and 9.5 for hydrogen-rich atmospheres, and between
$\Te=3500$~K and 40,000~K and $\logg=7.0$ and 9.0 for helium-rich
atmospheres. Note that we now include in our hydrogen models the
opacity from the red wing of Ly$\alpha$ calculated by \citet{KS2006}
and kindly provided to us by P. Kowalski, which is known to affect the
flux in the ultraviolet region of the energy distribution. The
calculations of absolute magnitudes also require the use of
mass-radius relations for white dwarfs, which are based on
evolutionary models similar to those described in \citet{fontaine01}
but with C/O cores, $q({\rm He})\equiv \log M_{\rm
  He}/M_{\star}=10^{-2}$ and $q({\rm H})=10^{-4}$, which are
representative of hydrogen-atmosphere white dwarfs, and $q({\rm
  He})=10^{-2}$ and $q({\rm H})=10^{-10}$, which are representative of a 
helium-atmosphere. From these color-magnitude relations at constant
$\logg$ values, we can obtain the corresponding relations at constant
{\it mass} values using the same evolutionary models as described
above.

We present below our results on the SDSS, GALEX, and 2MASS photometric
systems, as well as on the USNO-B1.0 photographic system. Synthetic
colors were calculated from the bandpasses available from the SDSS
website\footnote{http://www.sdss.org/dr6/instruments/imager/\#filters}
and discussed in \citet{Fukugita1996}, while the 2MASS filters are
described in \citet{Cohen2003} and the transmission functions were
taken from the survey
website\footnote{http://www.ipac.caltech.edu/2mass/releases/allsky/doc/ \\
        sec6\_4a.html}.
Similarly, GALEX synthetic colors were calculated from the bandpasses
available from the GALEX
website\footnote{http://galexgi.gsfc.nasa.gov/docs/galex/Documents/ \\
        PostLaunchResponseCurveData.html}
and described in \citet{Morrissey2004}, and information about the
filters for the USNO-B1.0 are given in \citet{monet03}, and
transmission curves are available from the Digitized Sky Survey
website\footnote{http://www3.cadc-ccda.hia-iha.nrc-cnrc.gc.ca/dss/}.

\subsubsection{$M_g$ vs $g-z$ Calibration}
\begin{figure}[h!]
\epsscale{1.5}
\centering
\plotone{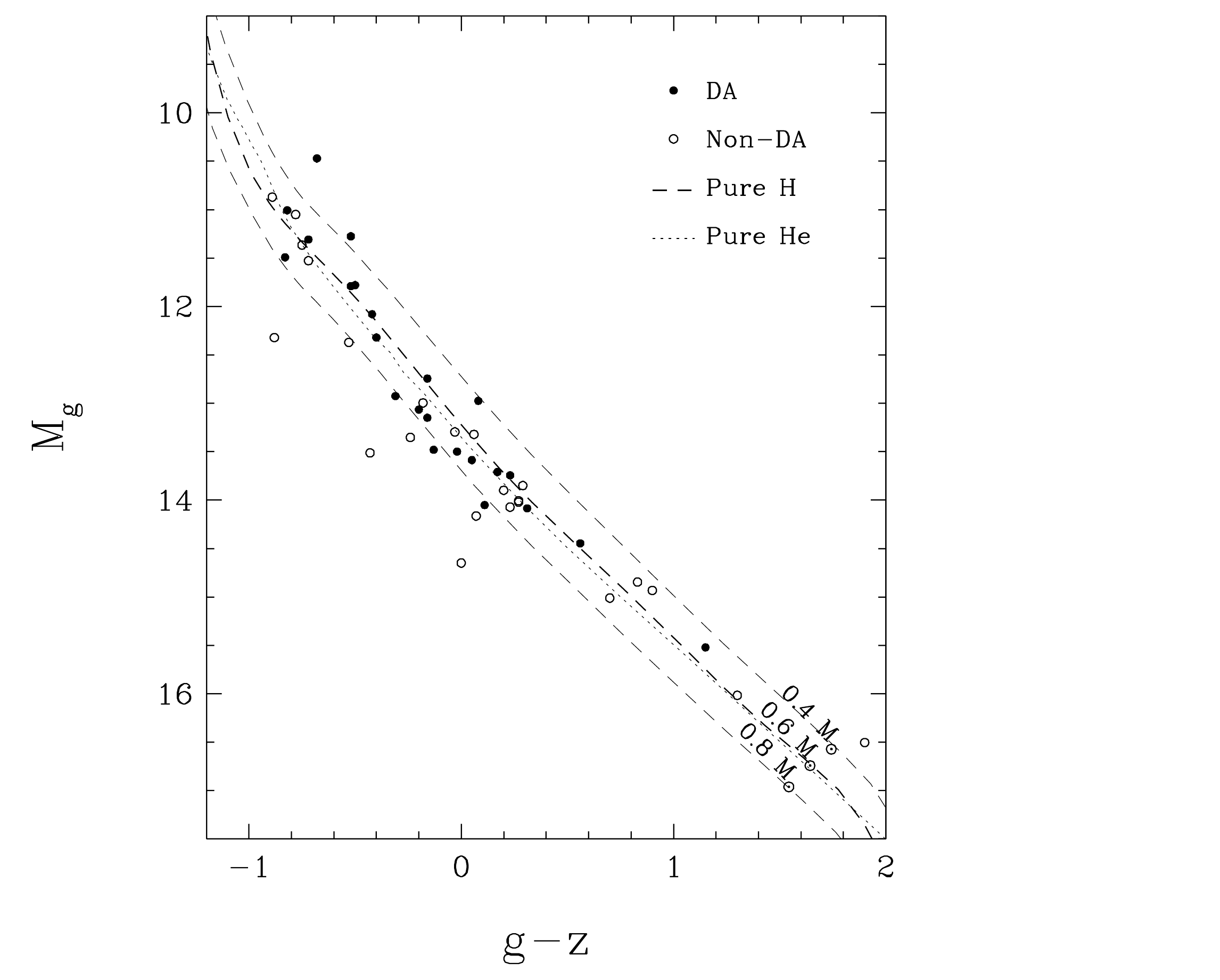}
\caption{Theoretical color-magnitude relations for pure
  hydrogen-atmosphere white dwarfs at 0.4, 0.6, and 0.8 \msun\ (dashed
  lines) and for pure helium-atmosphere white dwarfs at 0.6 \msun\
  (dotted line).  The pure hydrogen sequence at 0.6 \msun\ is used to
  determine absolute $M_g$ magnitudes for stars with SDSS photometry.
  Also shown are the 50 white dwarfs from the WD Catalog with
  available $ugriz$ photometry and trigonometric parallaxes from the
  Yale Parallax Catalog.\label{fig5}}
\end{figure}
We show in Figure \ref{fig5} the theoretical $M_g$ vs $g-z$
color-magnitude relation for hydrogen-atmosphere white
dwarfs at 0.6 \msun\ together with similar sequences at 0.4 \msun\ and 0.8 \msun,
which are representative of the intrinsic mass distribution for DA
stars (see, e.g., \citealt{gianninas2011}). These
are used below to evaluate the accuracy of our color-magnitude
calibration. Also shown in Figure \ref{fig5} is a single 0.6 \msun\
helium-atmosphere sequence used to evaluate the influence
of the unknown atmospheric composition on the
color-magnitude relations. As can be seen, the 0.6 \msun\ helium
sequence follows closely the corresponding hydrogen sequence in this
particular diagram, and it is thus perfectly justified to rely on the
hydrogen-rich sequence only to evaluate the photometric distances to our
objects.

As an external verification of our color-magnitude relations, we also
plot in Figure \ref{fig5} the 50 spectroscopically confirmed white
dwarfs from the WD Catalog that also have trigonometric parallax
measurements published in the Yale Parallax Catalog \citep{YPC1995}
and with SDSS photometry available; we consider here only stars with
parallax uncertainties less than 30\%. We also distinguish DA and
non-DA stars. The absolute magnitudes $M_g$ are directly obtained from
the distance modulus $M_g=g+5\log\pi+5$. Our results show that the
observed scatter with respect to the 0.6 \msun\ theoretical sequence
is entirely consistent with that expected from the intrinsic white
dwarf mass distribution, as indicated by the theoretical sequences at
0.4 and 0.8 \msun. This can be tested more quantitatively by measuring
the mass of each star directly from the color-magnitude diagram using
our theoretical sequences and a simple interpolation scheme. We obtain a
mean mass of $\langle M \rangle=0.63$ \msun, with a dispersion of
$\sigma(M)=0.20$ \msun, entirely consistent with the photometric mass
distribution obtained by \citet[][see their section 5.3]{bergeron01}
for the sample of cool white dwarfs with trigonometric parallaxes from
the Yale Parallax Catalog, $\langle M \rangle=0.65$ \msun,
$\sigma(M)=0.20$ \msun. As also discussed by Bergeron et al., however,
the dispersion expected from spectroscopic mass distributions are
considerably smaller, typically $\sigma(M)\sim 0.15$
\msun\ \citep[see, e.g.,][]{gianninas2011} due to the increased
sensitivity to $\logg$ of the spectroscopic technique over the
photometric method based on trigonometric parallax measurements. We
thus conclude from this comparison that our $M_g$ vs $g-z$
color-magnitude relation is well calibrated.

Finally, we use the $M_g$ vs $g-z$ theoretical relation for all white
dwarf candidates in our sample with observed $ugriz$ photometry to
estimate a photometric distance for each object assuming a
hydrogen-atmosphere and a mass of 0.6 \msun, the average mass for
white dwarfs. The photometric distances for 4823 white dwarf
candidates in our sample, identified from the $(H_g, g-z)$ reduced
proper motion diagram, are estimated in this manner.

\subsubsection{\mv\ vs \nuv\ Calibration}
\begin{figure}[h!]
\centering
\epsscale{1.5}
\plotone{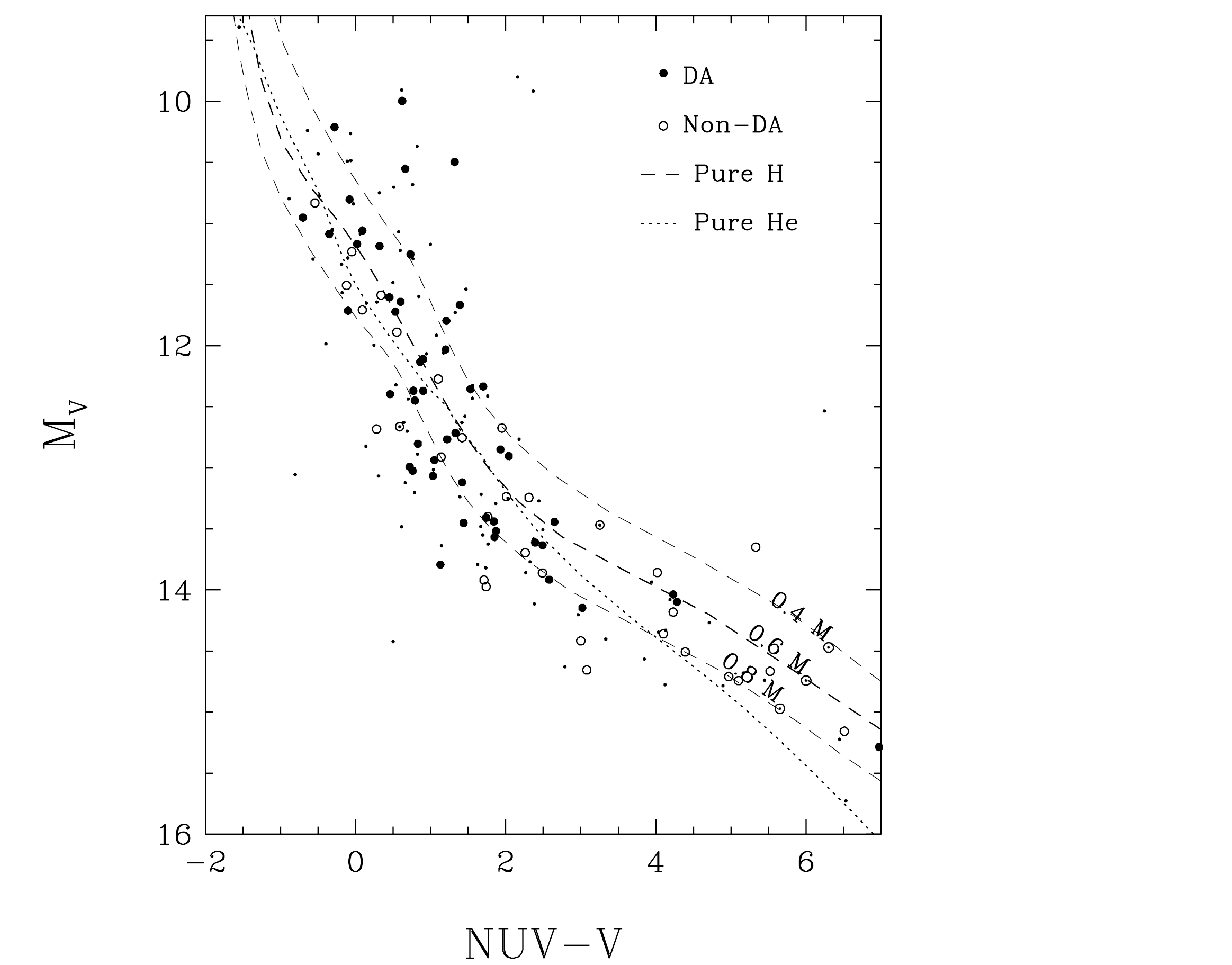}
\caption{Same as Figure \ref{fig5} but for the
  determination of absolute $M_V$ magnitudes for stars with GALEX
  photometry. Also shown are the 82 white dwarfs from the WD Catalog
  with available GALEX photometry and trigonometric parallaxes from
  the Yale Parallax Catalog. The small dots correspond to USNO photographic
  magnitudes while the larger symbols make use of Johnson $V$ magnitudes.
\label{fig6}}
\end{figure}

In an approach similar to that described in the previous section, we
use the theoretical relation between the absolute magnitude \mv\ and
the color index \nuv\ to determine an absolute magnitude for every
object with observed GALEX photometry, assuming a hydrogen atmosphere
at 0.6 \msun. We show in Figure \ref{fig6} the theoretical \mv\ vs
\nuv\ color-magnitude relations for the same mass values and
atmospheric compositions as above. In contrast with the results from the previous
section, the helium sequence starts to differ from the hydrogen
sequence at 0.6 \msun\ for \nuv\ $>2$, or $\Te<8000$~K, and becomes
significantly different for \nuv\ $>4$, or $\Te<6000$~K. However,
since we do not expect to identify many white dwarfs cooler than
$\sim6000$~K on the basis of their UV magnitudes, it is justified to
rely solely on the pure hydrogen sequence.

Figure \ref{fig6} also shows the 82 white dwarfs from the WD Catalog
observed by GALEX, with trigonometric parallax measurements available
in the Yale Parallax Catalog.  Since we are mostly interested here in
verifying the validity of our color-magnitude relations, we want to
use the best photometry available for each star in order to reduce the
scatter related to the uncertainty of photographic magnitudes. Hence,
in addition to $V$ magnitudes estimated from USNO photographic
magnitudes (small dots in Figure \ref{fig6}), we also show the results
obtained using apparent $V$ magnitudes measured on the Johnson
photometric system, and taken from the literature (larger symbols in
Figure \ref{fig6}). This comparison is thus analogous to that shown in
Figure \ref{fig5}, and we see once again that the bulk of white dwarfs
is well contained between the 0.4 \msun\ and 0.8 \msun\ theoretical
sequences. Performing the same calculations as before, we find a mean
mass of $\langle M \rangle=0.67$ \msun\ and a dispersion of
$\sigma(M)=0.22$ \msun, still consistent with the values obtained
by \citet{bergeron01}.

We finally apply this calibration to all white dwarf candidates
identified from the $(H_V,\rm{NUV}-V)$ reduced proper motion diagram,
and estimate photometric distances for 8092 stars with GALEX
photometry (but with no SDSS counterparts). 

\begin{figure}[h!]
\epsscale{1.5}
\plotone{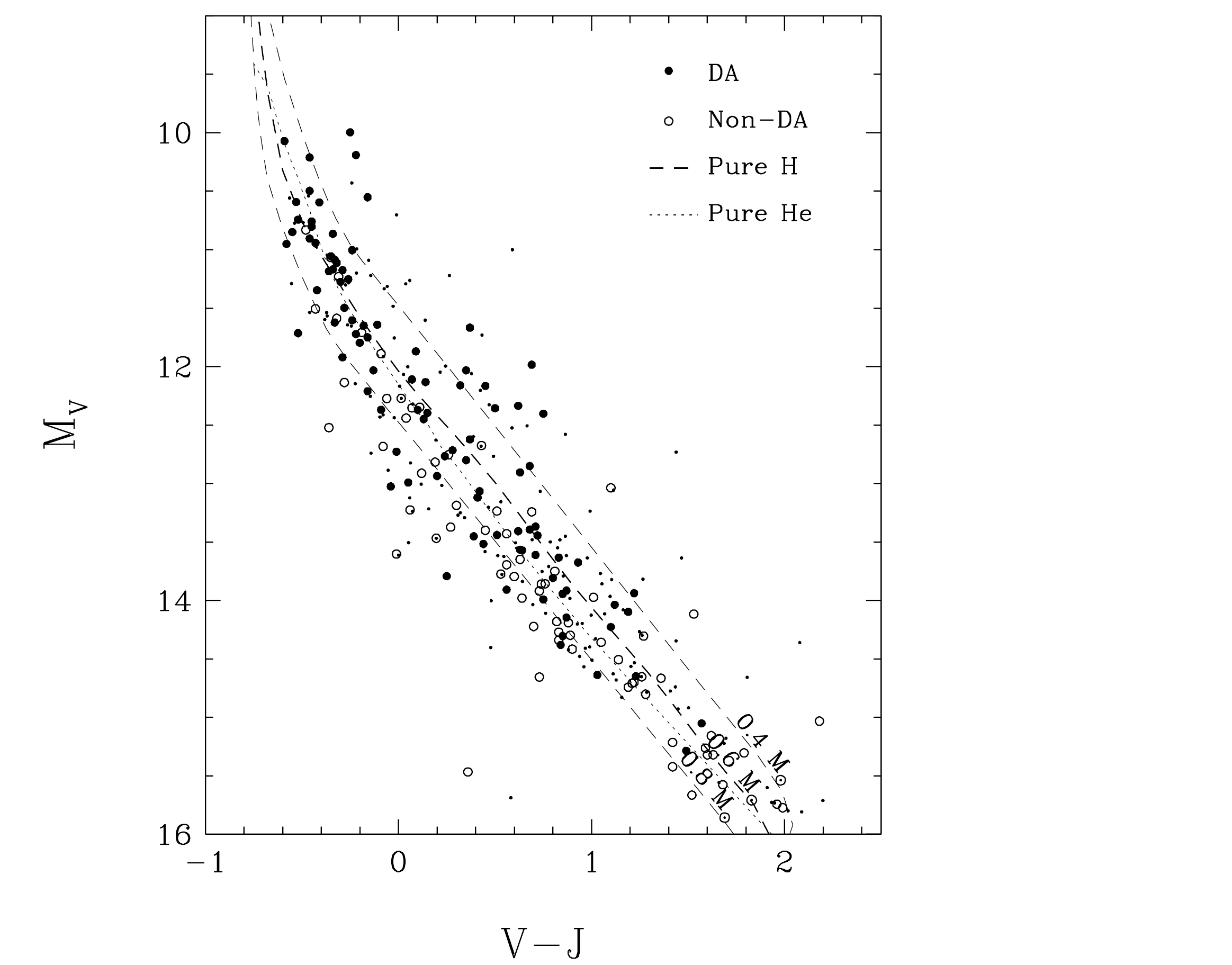}
\caption{Same as Figure \ref{fig5} but for the
  determination of absolute $M_V$ magnitudes for stars with 2MASS
  photometry.  Also shown are the 167 white dwarfs from the WD Catalog
  with available 2MASS photometry and trigonometric parallaxes from
  the Yale Parallax Catalog. The small dots correspond to USNO photographic
  magnitudes while the larger symbols make use of Johnson $V$ magnitudes.
  \label{fig7}}
\end{figure}

\subsubsection{\mv\ vs $V-J$ Calibration}

The theoretical relations between the absolute \mv\ magnitude and
color index $V-J$ are shown in Figure \ref{fig7}. Also displayed are
the 167 white dwarfs from the WD Catalog for which a trigonometric
parallax measurement in the Yale Parallax Catalog and a $V-J$ color
index are available (both Johnson $V$ and USNO photographic magnitudes
are displayed, as explained in the previous section). 

We note again in this figure that most of the points are contained
between the 0.4 \msun\ and 0.8 \msun\ theoretical sequences, with
$\langle M \rangle=0.67$ \msun\ and $\sigma(M)=0.21$ \msun, and that
the hydrogen- and helium-atmosphere sequences agree sufficiently
enough to assume a hydrogen-rich composition for all objects in our
sample. We obtain the absolute magnitudes and photometric distances
for all 1132 white dwarf candidates having counterparts in the 2MASS
catalog (but with no SDSS or GALEX counterparts), assuming a
hydrogen-atmosphere and a mass of 0.6 \msun.

\subsubsection{$M_V$ vs $V-I_N$ Calibration}
\begin{figure}[h!]
\epsscale{1.5}
\plotone{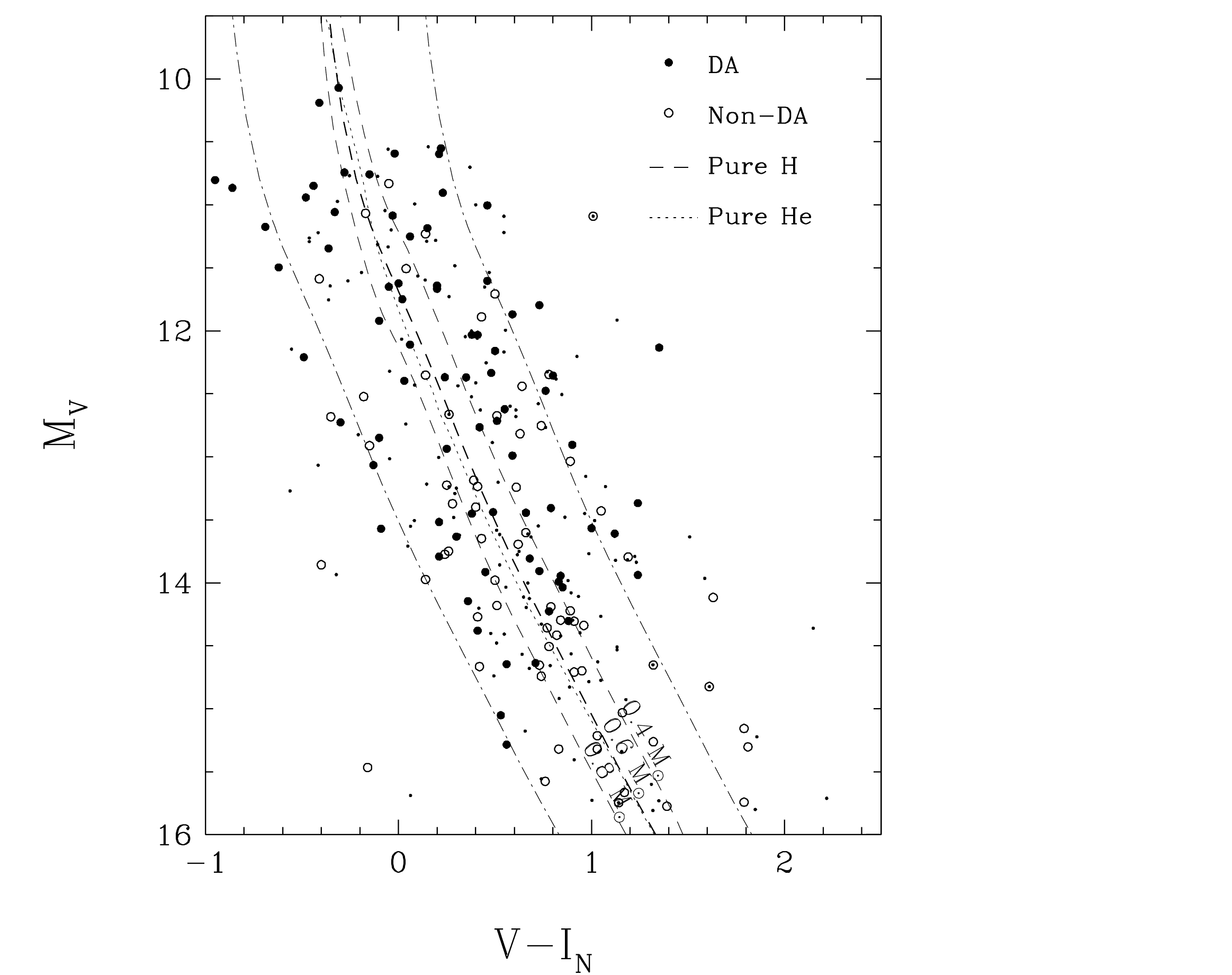}
\caption{Same as Figure \ref{fig5} but for the
  determination of absolute $M_V$ magnitudes for stars with USNO-B1.0
  photographic magnitudes.  Also shown are the 151 white dwarfs from
  the WD Catalog with available USNO photographic magnitudes and
  trigonometric parallaxes from the Yale Parallax Catalog; the small
  dots correspond to USNO photographic magnitudes while the larger
  symbols make use of Johnson $V$ magnitudes. The dot-dashed lines
  represent the 0.5 mag dispersion around the pure hydrogen sequence
  at 0.6 \msun\ estimated from the accuracy of the $B_J$, $R_F$, and
  $I_N$ magnitudes.\label{fig8}}
\end{figure}
Figure \ref{fig8} displays our last color-magnitude relation, that
between \mv\ and the photographic color index $V-I_N$. Also shown are
the 151 white dwarfs from the WD Catalog with trigonometric parallax
measurements in the Yale Parallax Catalog and $V-I_N$ color indices
available. Again, both Johnson $V$ and USNO photographic magnitudes
are used in this plot. Not unexpectedly, the comparison between the
observed \mv\ values and the theoretical sequences reveals a much
larger scatter, larger than that expected from the intrinsic mass
distribution alone. Indeed, we measure a mean mass of $\langle M
\rangle=0.58$ \msun, which is 0.07 \msun\ lower than the value of
\citet{bergeron01} for a similar sample, but more importantly, we find
a significantly larger dispersion of $\sigma(M)=0.26$ \msun, compared
to the value of 0.20 \msun\ obtained by Bergeron et al. This is a direct
consequence of the lesser accuracy of photographic magnitudes. Indeed,
most of the scatter observed here is likely due to the 0.5 mag
uncertainty in photographic $I_N$ magnitudes \citep{monet03}. For
instance, we also illustrate in Figure \ref{fig8} the effect of a 0.5
mag error on the $V-I_N$ color-index for the theoretical hydrogen
sequence at 0.6 \msun. Most of the points are then contained within
these boundaries. The reliability of the color-magnitude relation is
therefore limited by the accuracy of the photographic
magnitudes. Unfortunately, these photographic magnitudes are the only
information available to estimate photometric distances for the 6688
candidates with no SDSS, GALEX, or 2MASS counterparts.

We finally conclude from the figures above that the color-magnitude
relations derived from SDSS, GALEX, and 2MASS photometry are
comparable in their level of accuracy, and that the least accurate 
photometric distances, as one might expect, are those estimated from photographic 
magnitudes.

\subsection{Error on Photometric Distances}
 
There are 70 spectroscopically confirmed white dwarfs with measured
trigonometric parallaxes recovered by our four reduced proper motion
diagrams. For these 70 white dwarfs, we can estimate photometric
distances using the photometric system corresponding to the reduced
proper motion diagram where each object was identified. For instance,
19 objects were identified on the basis of their SDSS photometry, that
is, they were selected from the $(H_g,g-z)$ reduced proper motion
diagram while their photometric distance was estimated using the $M_g$
vs $(g-z)$ color-magnitude calibration.  Similarly, 8 objects were
identified with the help of GALEX photometry, 35 from 2MASS
photometry, and 8 from photographic magnitudes. We finally end up with
a sample of 70 confirmed white dwarfs with distances measured from
trigonometric parallax measurements, where each star has also been
identified in at least one of the reduced proper motion diagrams, and
with a corresponding photometric distance estimate.

\begin{figure}[h!]
\epsscale{1.3}
\plotone{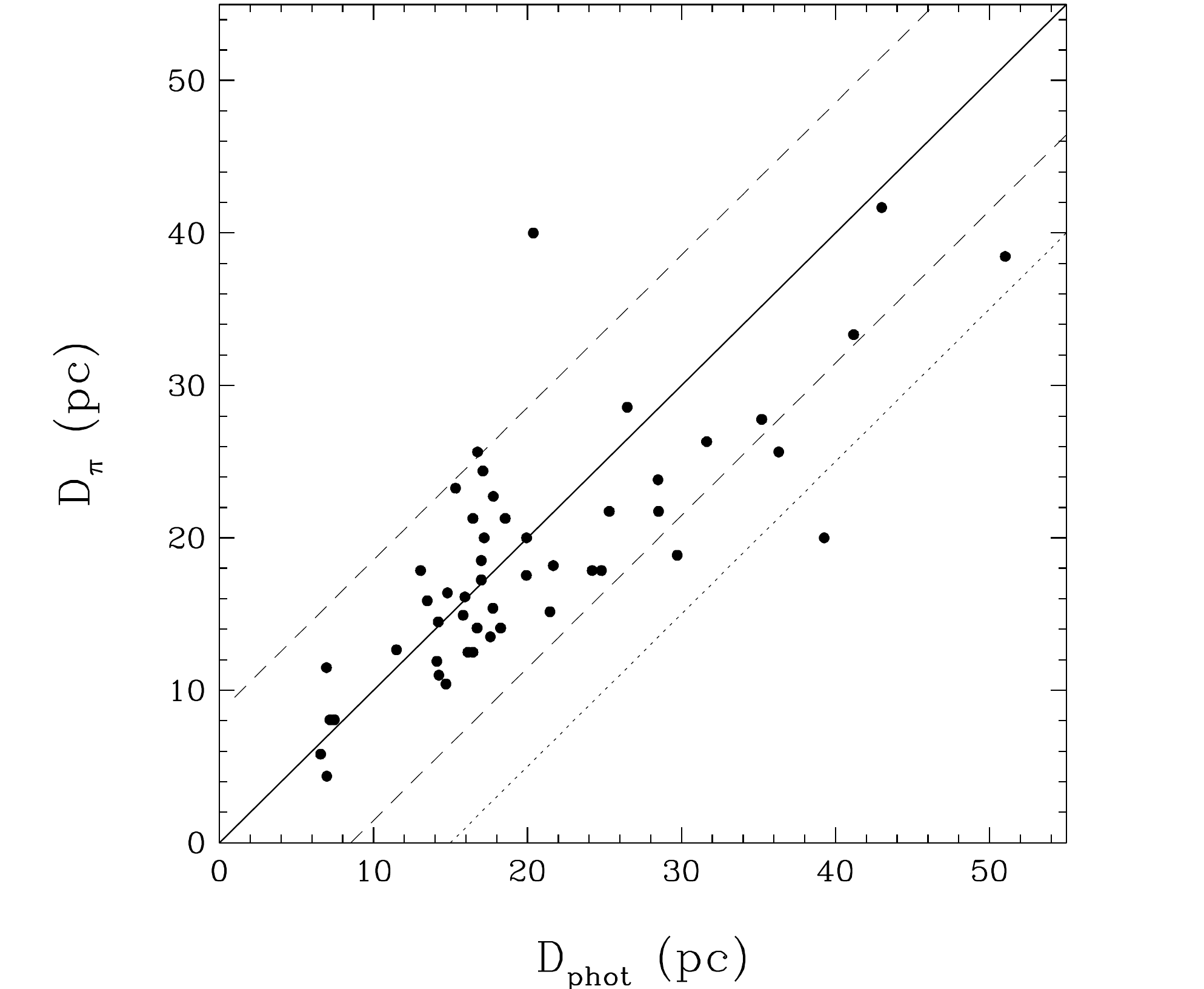}
\caption{Distances obtained from trigonometric
  parallaxes compared with photometric distances estimated from
  theoretical color-magnitude relations. The dots represent the 70
  spectroscopically confirmed white dwarfs with parallax uncertainties
  less than 30\% and also selected from our reduced proper motion
  diagrams. The solid line represents the 1:1 relation, while the
  dashed lines represent the $1\sigma$ dispersion of $8.5$ pc
  resulting from the combined errors of the 4 color-magnitude
  relations. The dotted line indicates the $+15$ pc error adopted in
  our analysis to ensure all white dwarf candidates within 40 parsecs
  of the Sun are included in our sample. \label{fig9}}
\end{figure}
The distances obtained from parallax measurements are compared to
photometric distances in Figure \ref{fig9}. The dashed lines represent
the average 8.5 pc (rms) dispersion relative to the 1:1 relation
estimated using the white dwarfs displayed in Figures \ref{fig5} to
\ref{fig8} for all four sets of color-magnitude relations. Given
the observed dispersion in Figure \ref{fig9} and the fact that this
dispersion appears to increase with distance, we choose to include in
our list of white dwarf candidates within 40 parsecs of the Sun all
objects with an estimated photometric distance less than 55 pc. We
believe that this conservative buffer of 15 pc (dotted line in Figure
\ref{fig9}) is enough to include all white dwarfs that could
potentially lie within 40 pc of the Sun. Of course, we keep in mind
that the main purpose of this calibration is purely to identify the
nearest objects. Subsequent spectroscopic analyses are
expected to provide more accurate distances, and lead to an
independent estimation of the error on the photometric distances.
 
\subsection{List of White Dwarf Candidates}

We are now able to determine photometric distances for each object on 
our list of white dwarf candidates following the method described in
the previous sections. If the estimated distance places it within 55
parsecs of the Sun, the object becomes part of our list of candidates
for follow-up spectroscopy. This process leaves us with a list of 1978
spectroscopic targets. The sample can be further reduced by
eliminating all previously known white dwarfs. To do so, the
coordinates of each candidate are compared to those listed in the WD
Catalog, and then with every star in the Simbad Astronomical
Database\footnote{http://simweb.u-strasbg.fr/} within a 1 arcmin
radius. This way, we found 499 white dwarfs in our candidate list that
were previously known, and an additional 35 objects with a known
spectral type that identifies them either as a main sequence stars or a 
background galaxy. The presence of these 9 galaxies in our candidate list 
indicates that there are apparently some spurious objects with false proper 
motions in the SUPERBLINK catalog. From all these objects with a known spectral
type, we can evaluate the contamination of our sample of white dwarf
candidates to be $\sim8\%$. 

With the objects with a known spectral type removed from our candidate
list, we finally end up with a list of 1341 targets for follow-up
spectroscopy.  This sample divides into 268 candidates identified on
the basis of SDSS photometry, 130 from GALEX photometry, 731 from
2MASS photometry, and 212 from USNO-B1.0 photographic magnitudes. The
candidates identified with SDSS and GALEX colors are given first
priority for follow-up spectroscopy, followed by stars with 2MASS
photometry. Finally, objects with only photographic magnitudes
available are given the lowest priority. 

\begin{figure}[h!]
\epsscale{1.3}
\plotone{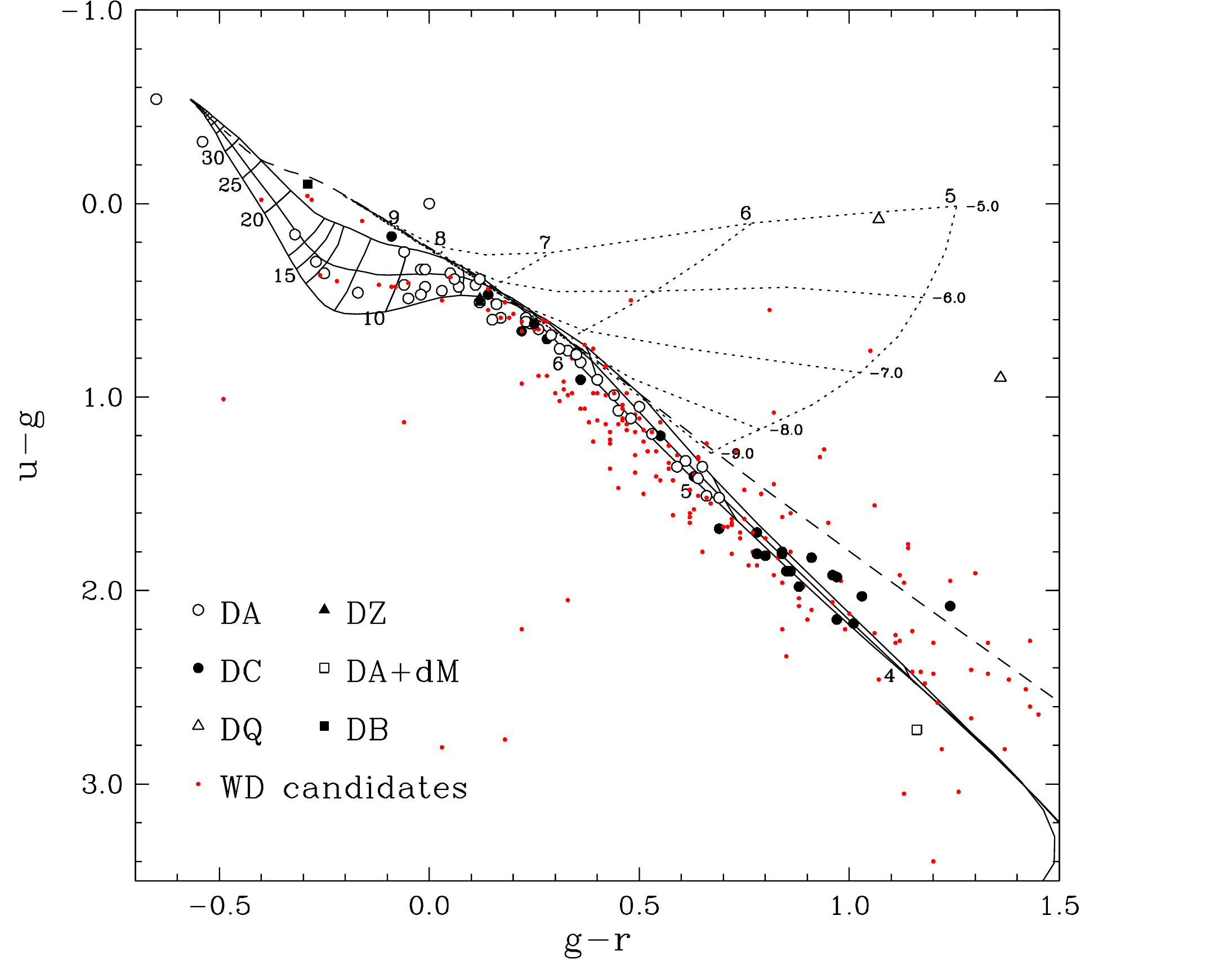}
\caption{($u-g$, $g-r$)
  color-color diagram showing the 268 white dwarf candidates with
  $ugriz$ photometry available. The 76 white dwarfs spectroscopically
  confirmed in Section 5 are shown with various symbols explained in
  the legend, while those without spectroscopic data are shown with red
  dots. The solid curves represent pure hydrogen model atmospheres at
  $\log g=7.0$, 8.0, and 9.0 (from bottom to top); effective
  temperatures are indicated in units of $10^3$ K. The dashed curve
  corresponds to pure helium atmospheres at $\log g=8.0$, and the dotted
  lines represent DQ models for 5 different
  compositions, from $\log \rm{C/He}=-9.0$ to $-5.0$. \\ \label{fig10}}
\end{figure}

The 268 white dwarf candidates with available SDSS photometry are
shown in Figure \ref{fig10} in a ($u-g$, $g-r$) color-color
diagram, together with theoretical predictions for our pure hydrogen,
pure helium, and DQ models.  This two-color diagram reveals that our
sample is dominated by cool white dwarfs. We also expect that an
important fraction of these cool objects will have a
hydrogen-dominated atmosphere, and that some candidates are most
likely DQ stars.

\begin{figure}[h!]
\epsscale{1.3}
\plotone{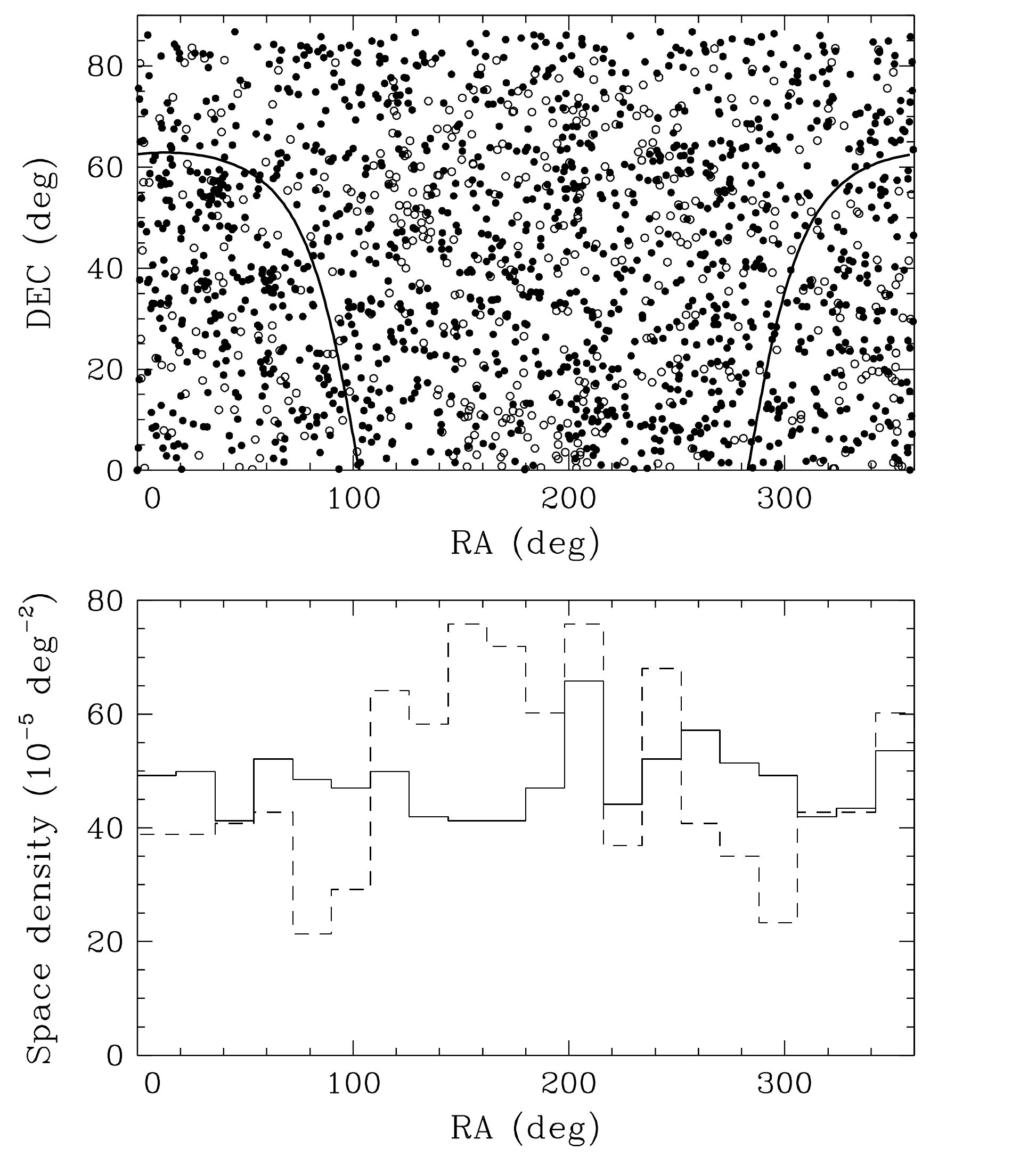}
\caption{Upper panel: Equal cylindrical projection
  of the equatorial coordinates for the sample of 1341 white dwarf
  candidates identified from SUPERBLINK (solid circles) compared with
  the sample of 499 stars from the WD Catalog recovered by our
  selection criteria (open circles). Also shown by the bold solid line
  is the region of the galactic plane. Lower panel: Space density as a
  function of right ascension, normalized to the total number of stars
  in each bin (both lines are thus on a comparable scale). The solid
  line represents the 1341 white dwarf candidates, while the dashed
  line corresponds to the white dwarfs from the WD
  Catalog. \label{fig11}}
\end{figure}

The 1341 white dwarf candidates identified in SUPERBLINK are also
displayed in Figure \ref{fig11}. The upper panel shows their
space distribution using a cylindrical equal-area projection of the
equatorial coordinates, while in the lower panel, the same
distribution is binned into a histogram to illustrate more clearly the
space density as a function of right ascension. This figure shows that
our white dwarf candidates are distributed uniformly on the sky,
without any drop in surface density near the galactic plane. The
spectroscopically confirmed white dwarfs from the WD Catalog are also
shown for comparison. Given this characteristic of our survey, we hope
to increase significantly the completeness of the local white dwarf
sample in this region, which has often been neglected in the past 
due to field crowding near the Galactic plane.

\section{Spectroscopic follow-up of the candidates}

We provide in this section a status report of our ongoing spectroscopic survey,
and present results for a first set of 422 objects from our target list, or nearly
a third of our complete sample. 
\begin{deluxetable*}{lllcccc}
\tabletypesize{\scriptsize}
\tablecolumns{12}
\tablewidth{0pt}
\tablecaption{Spectroscopic Observing Runs}
\tablehead{
\colhead{} &
\colhead{} &
\colhead{} &
\colhead{Grating} &
\colhead{Blaze} &
\colhead{Coverage}&
\colhead{Slit}\\
\colhead{Date} &
\colhead{Telescope} &
\colhead{Spectrograph} &
\colhead{(l mm$^{-1}$)} &
\colhead{(\AA)} &
\colhead{(\AA)}&
\colhead{($\arcsec$)}}
\startdata
2009 May&Steward Observatory Bok 2.3 m&B$\&$C&600	&3568&3800-5600&4.5\\
2009 August&NOAO Mayall 4 m&RC&527 &5540&3800-6800&2\\
2009 November&Steward Observatory Bok 2.3 m&B$\&$C&600&3568&3800-5600&4.5\\
2009 December&NOAO 2.1 m&Goldcam&600 &4900&3800-6700&2\\
2010 March&NOAO Mayall 4 m&RC&316 &4000&3900-6700&2\\
2010 May&NOAO 2.1 m&Goldcam&500 &5500&3800-6700&2\\
2010 July&Steward Observatory Bok 2.3 m&B$\&$C&400&4800&3800-6700&4.5\\
2010 October&NOAO Mayall 4 m&RC&316 &5500&3900-6700&2\\
\enddata
\label{tab2}
\end{deluxetable*}
\subsection{Spectroscopic Observations}

Optical spectra have been obtained with the Steward Observatory 2.3-m
telescope, and the NOAO Mayall 4-m and 2.1-m telescopes, during 8
different observing runs between 2009 May and 2010 October. The
adopted configurations allow a spectral coverage of
$\lambda\lambda$3200--5300 and $\lambda\lambda$3800--6700, at an intermediate
resolution of $\sim 6$~\AA\ FWHM.  Spectra were first obtained at low
signal-to-noise ratio (S/N $\sim25$), which is sufficient to identify
main sequence objects, but also represents the lower limit required
to obtain reliable model fits to the spectral lines. Table \ref{tab2}
summarizes our spectroscopic observing runs and observational
setups.

As a result of our spectroscopic observations, 193 newly identified
white dwarfs have been spectroscopically confirmed.  Among these, 68
were identified on the basis of SDSS photometry, 18 from GALEX, 70 from
2MASS, and 12 from USNO photographic magnitudes. The
remaining 25 objects were
discovered using an earlier version of our selection method, based on
criteria different from those adopted in the final version.  These
objects have revised photometric distance estimates beyond 55
pc, and are thus not included in our final list of 1341 candidates. \\
\\
\subsection{Spectroscopic Content}

Our spectroscopic follow-up observations identify 193 new white dwarfs, 
listed in Tables \ref{tab3} and \ref{tab4}. Table
\ref{tab3} provides astrometric data as well as NLTT and SDSS
designations, when available, while Table \ref{tab4} lists the same
objects but with the available photometry and adopted spectral
types. In the earlier presentation of our results \citep{lim10}, the
`LSPM J' notation was used for the object names, as in \citet{lspm05},
but the the proper motion limit for stars in SUPERBLINK has lowered to
$\mu>0.04"\rm{yr}^{-1}$, and the entries were homogenized with a `PM
I' designation. In summary, this subsample contains 127 DA (among
which 9 DA+dM and 4 magnetic), 1 DB, 56 DC, 3 DQ, and 6 DZ white
dwarfs.

Once we have a confirmed white dwarf, it is possible to
improve upon our preliminary distance estimates, which were based on
approximate $V$ magnitudes and color-magnitude relations, by making
use of the full set of photometric data.  Here we rely on the
so-called photometric method described at length in \citet[][and
references therein]{noemi2012} where the available photometry
for each star is fitted with theoretical fluxes,
properly averaged over each bandpass.  Both $\Te$ and the solid angle
$\pi (R/D)^2$ are considered free parameters, where $R$ is the radius
of the star and $D$ its distance from Earth.  We assume a value of
$\log g=8$ and corresponding radius $R$, and obtain directly the
distance $D$. These improved photometric distances are given in Table
\ref{tab4}.

When this survey was undertaken in 2009 May, none of these objects had
a white dwarf classification or a spectral type available in the
literature. But since then, white dwarf identifications have been
reported in \citet{kilic2010}, \citet{vennes2011}, \citet{Tonry2012},
and \citet{sayres2012}. We identified these stars in Table \ref{tab3}
and chose to leave these objects in our sample since they have been
discovered independently.  We also want to point out that even though
Table \ref{tab3} contains 25 stars with existing SDSS spectra, all our
targets have been identified using our own spectroscopic data.
 
\begin{figure*}[t!]
\epsscale{0.8}
\plotone{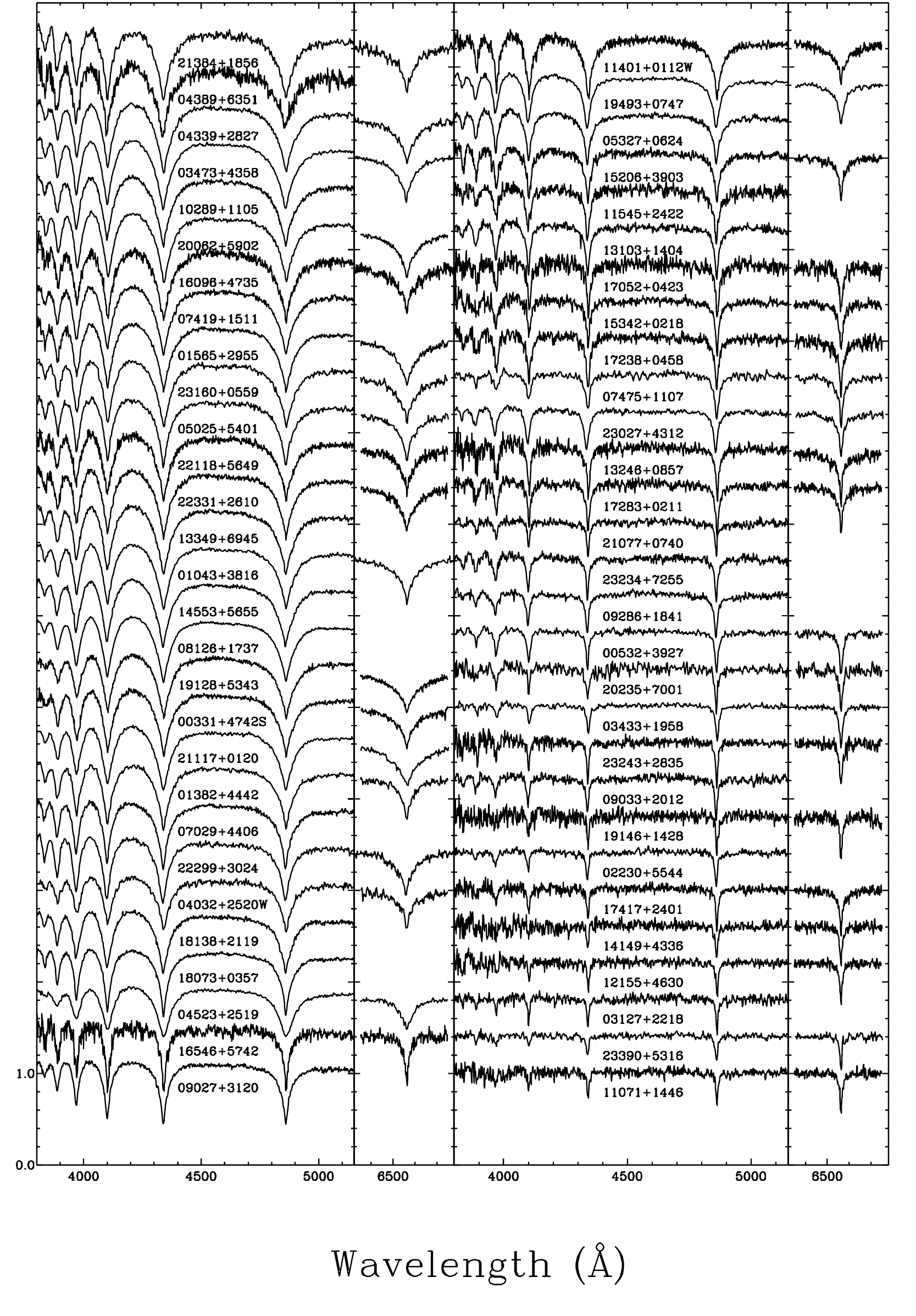}
\caption{(a) - Optical spectra for our sample of DA
  white dwarfs from SUPERBLINK. The spectra are displayed in order of
  decreasing equivalent width of \hb, from upper left to bottom right,
  and shifted vertically for clarity. The \ha\ line is also shown,
  when available, and normalized to a continuum set to
  unity.\label{DA}}
\end{figure*}

\addtocounter{figure}{-1}
\begin{figure*}
\epsscale{0.8}
\plotone{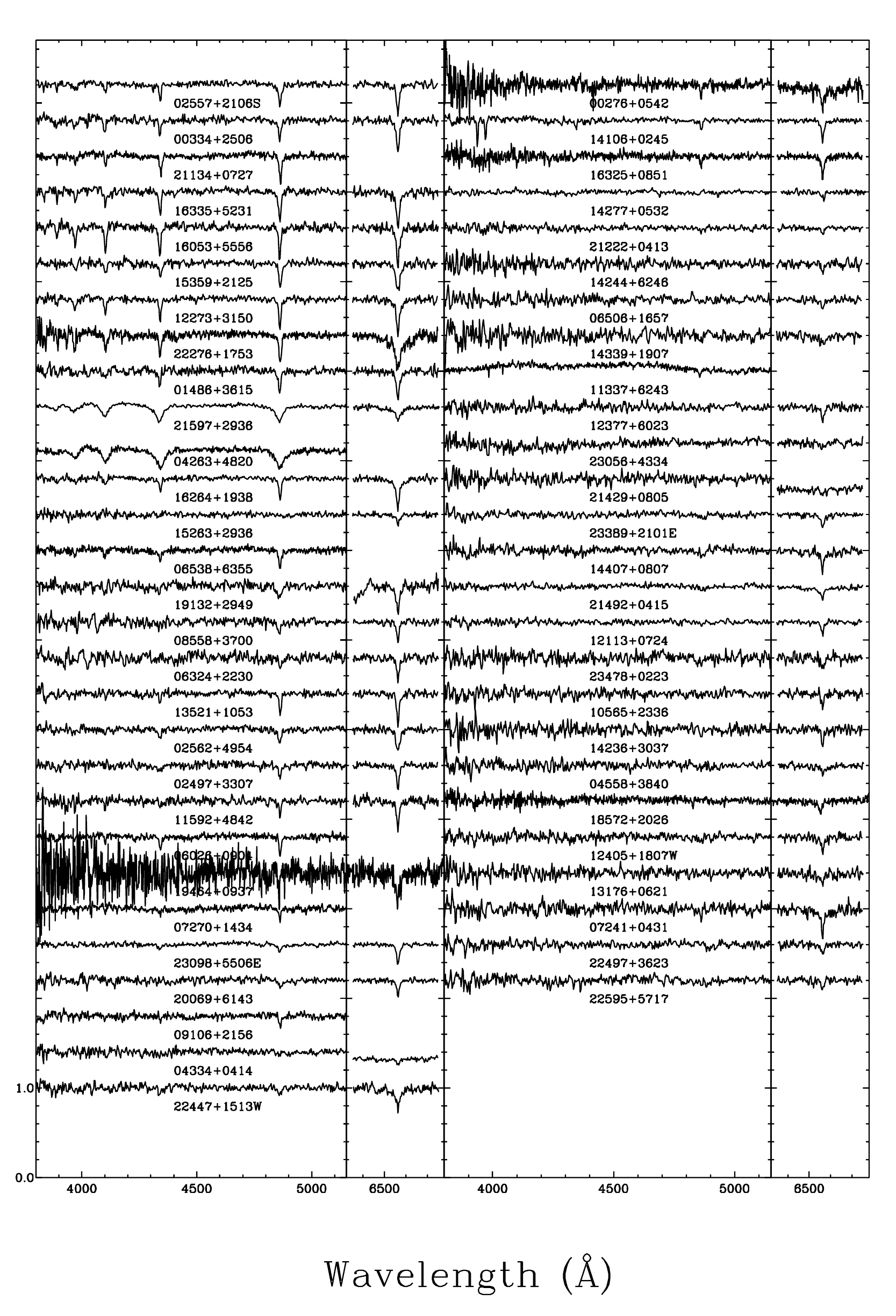}
\caption{(b) - continued.}
\end{figure*}

\begin{figure}[h!]
\plotone{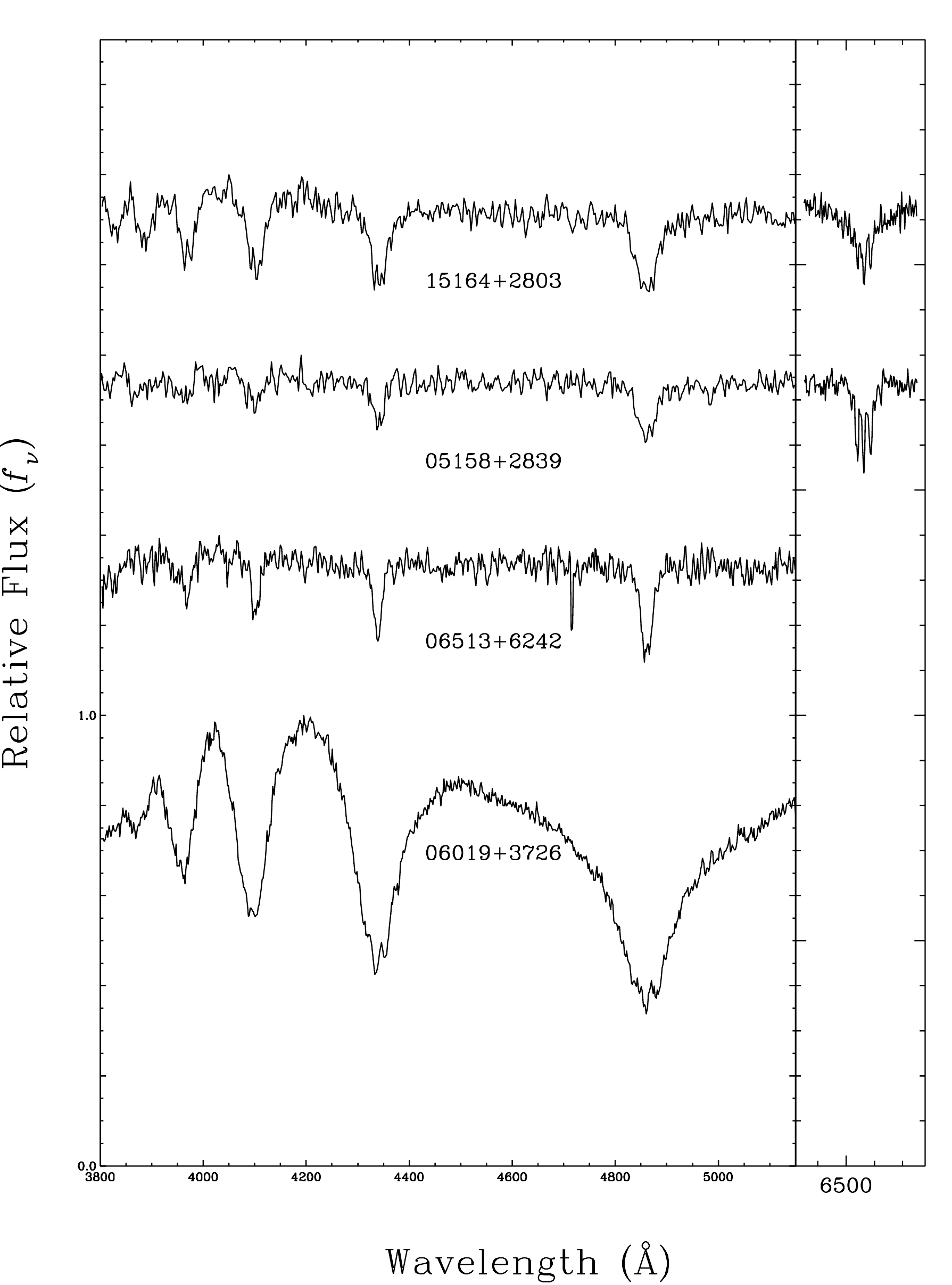}
\epsscale{1.0}
\caption{Our subsample of magnetic DA white dwarfs,
 shifted vertically for clarity. \label{fig13}}
\end{figure}
\begin{figure}[h!]
\epsscale{1.2}
\plotone{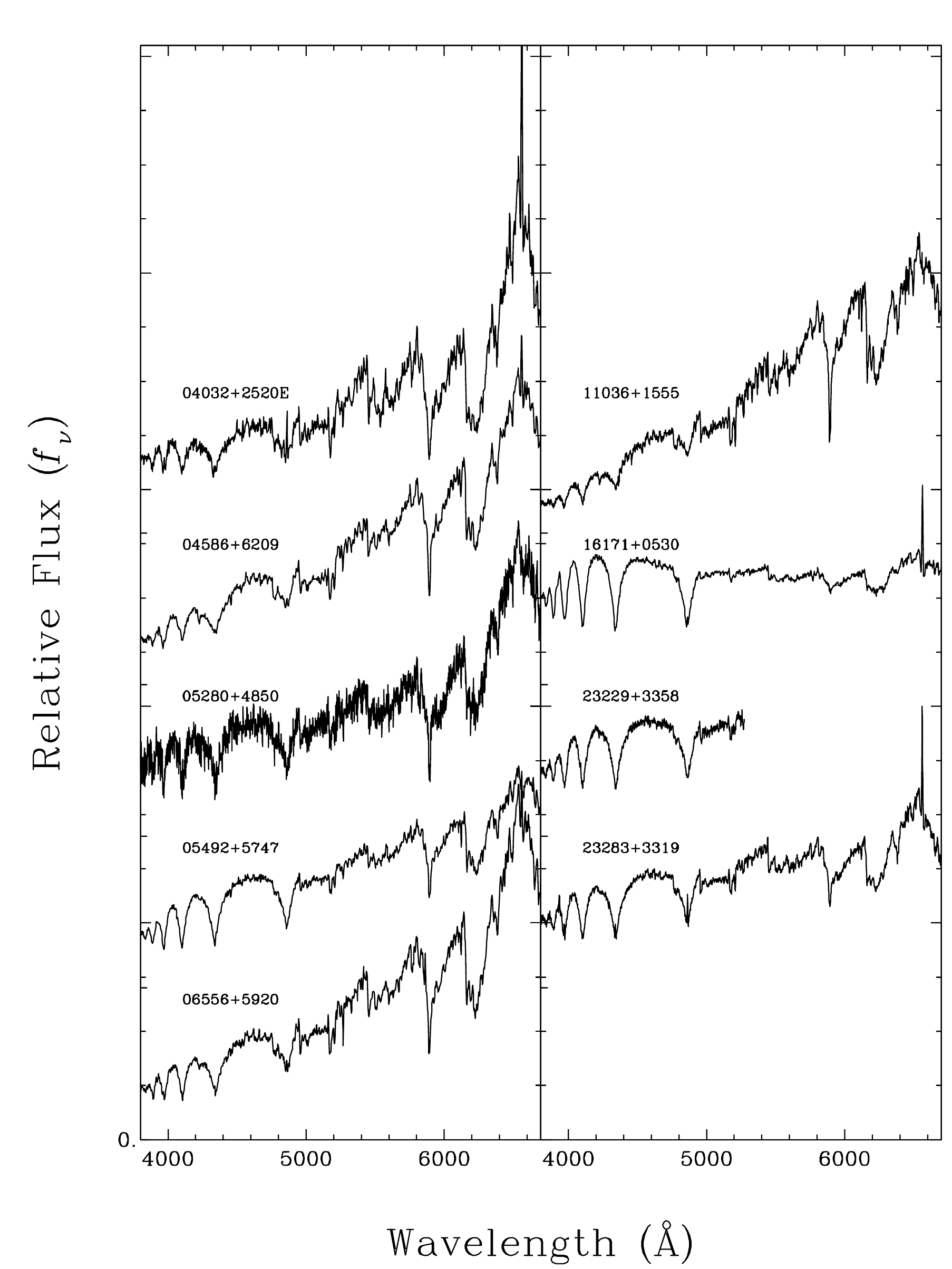}
\caption{Our sample of binary systems
  composed of a DA white dwarf and a M dwarf companion.  The spectra
  are shifted vertically for clarity. The \ha\ and \hb\ line cores of
  04032+2520E (not to be confused with the DA 04032+2520W), 06556+5920,
  16171+0530, and 23283+3319 are contaminated by emission from the M
  dwarf.\label{fig14}}
\end{figure}

Optical spectra from our subsample of DA stars covering $\rm{H}8$ to
\hb\ --- or \ha\ when available --- are shown in Figure \ref{DA}. Note
that 14106+0245 (right panel of Figure \ref{DA}b, second object from
the top) is a DAZ, and that our subsample of 4 magnetic DA white
dwarfs (05158+2839, 06019+3726, 06513+6242, and 15164+2803) are
displayed in Figure \ref{fig13}. Our survey also detected 9 new DA white
dwarfs with an M dwarf companion; these are plotted separately in
Figure \ref{fig14}. This was quite unexpected, since the cuts in the
reduced proper motion diagrams were chosen in order to avoid main
sequence stars. As a consequence, we avoided all objects that are
bright in the infrared portion of the spectrum. As discussed earlier,
however, this is true for our criteria in $g-z$, $V-J$, and $V-I_N$,
but not for our criteria in \nuv, which is efficient for detecting
blue objects. And indeed, all 9 DA+dM systems were detected using this
last reduced proper motion diagram.

\begin{figure}[h!]
\epsscale{1.3}
\plotone{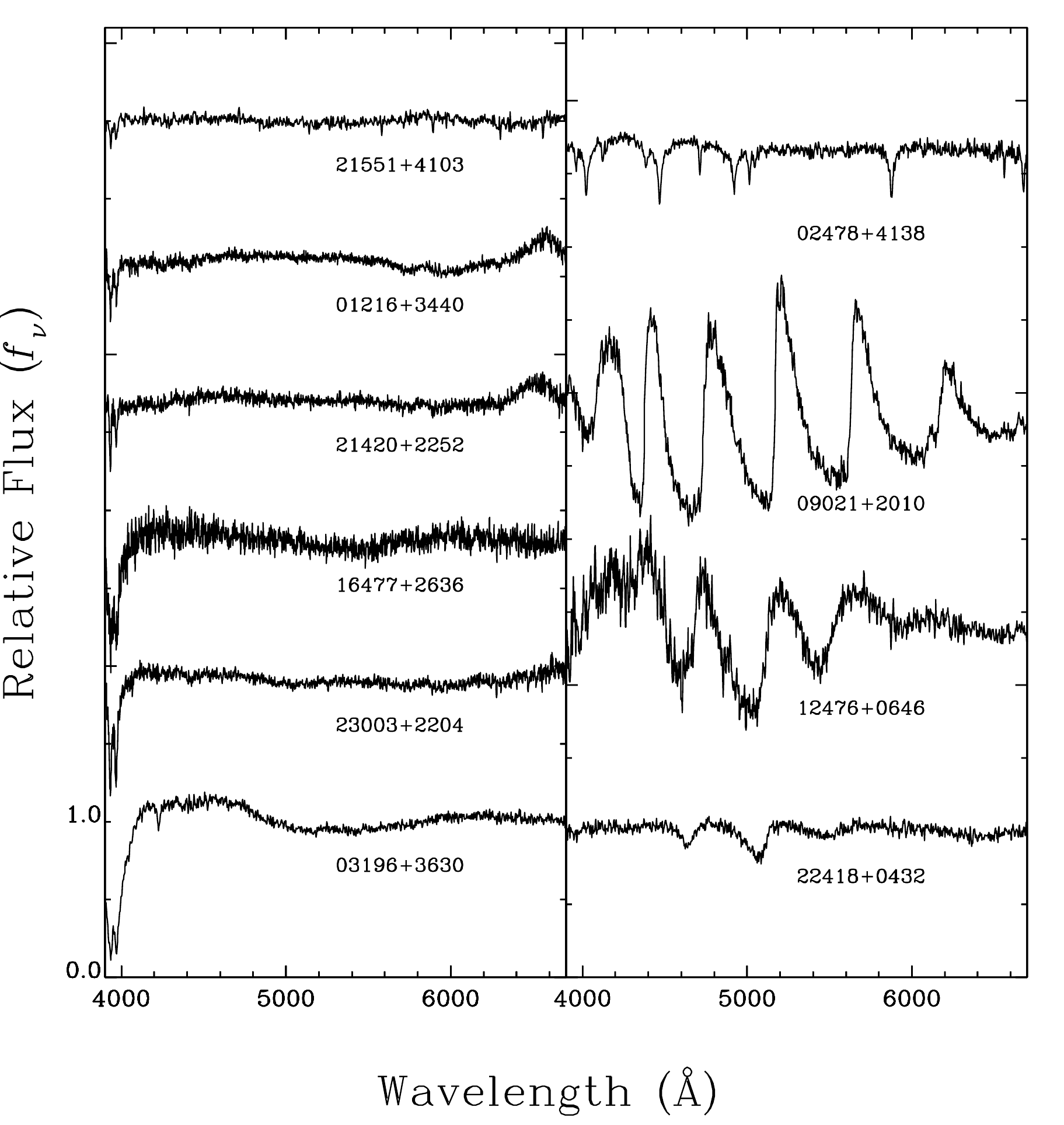}
\caption{Spectra of white dwarfs in our sample
  with helium-rich atmospheres. The panel on the left displays the DZ
  stars, while the panel on the right shows our only DB star, as well
  as 3 DQ stars. \label{fig15}}
\end{figure}

\begin{figure*}[t!]
\epsscale{0.7}
\plotone{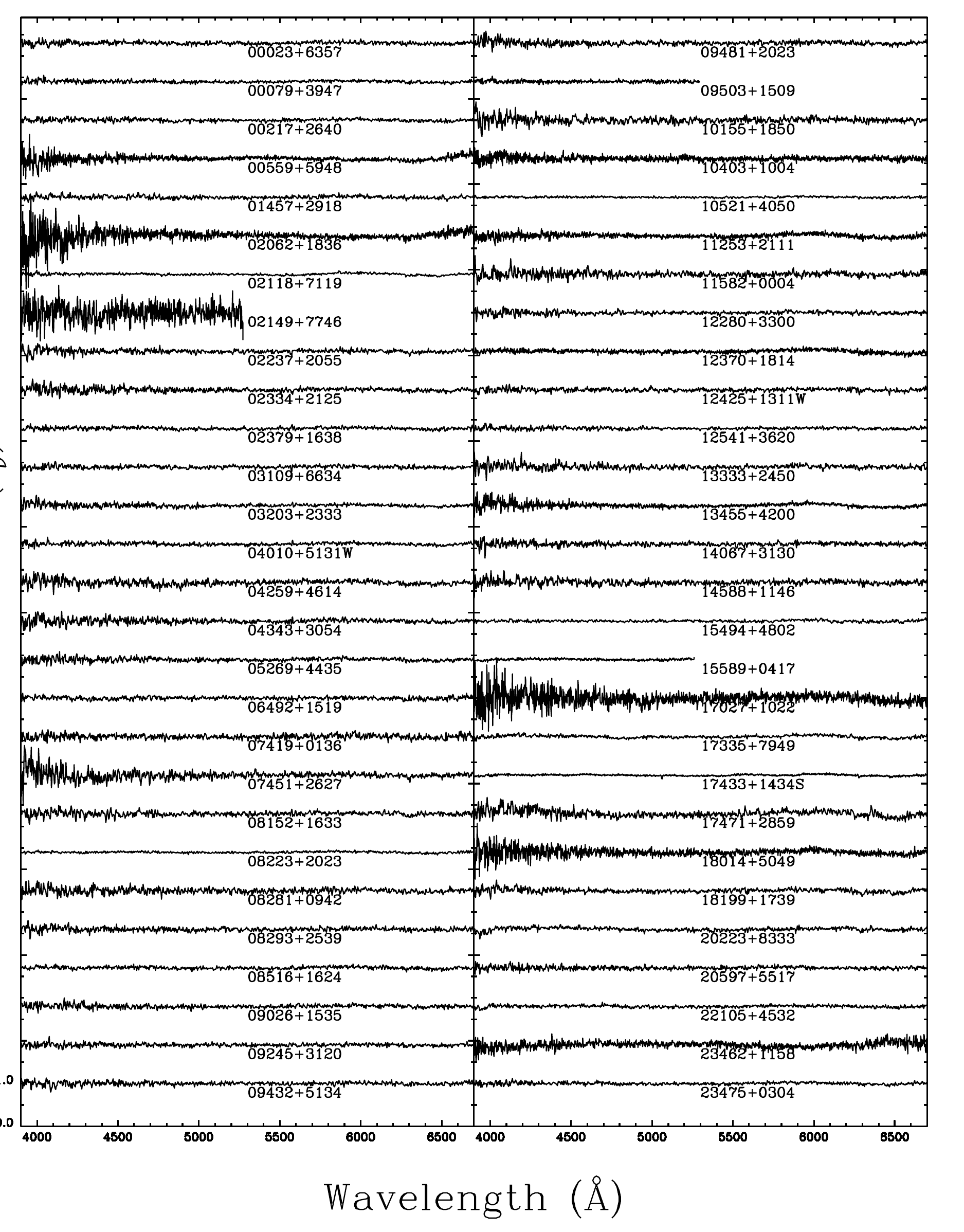}
\caption{Our sample of featureless DC stars. All
  spectra are normalized to a continuum set to unity and are offset
  from each other by a factor of 0.9. \label{fig16}}
\end{figure*}

The DB, DQ, and DZ stars in our sample are plotted together in Figure
\ref{fig15}. Unfortunately, the observational setup with the NOAO
telescopes does not allow the coverage of wavelengths shorter than
$\sim3900$ \AA, while covering \ha\ simultaneously. Calcium lines can
still be easily identified, however, but additional spectroscopic
observations near the $\sim3700$ \AA\ region are currently being
obtained in order to perform a proper model atmosphere analysis of
these DZ stars. In the case of DQ white dwarfs, one particular object,
12476+0646, exhibits the pressure-shifted carbon lines characteristic
of DQpec stars \citep{kowalski2010}.
Finally, we display our featureless DC spectra in Figure \ref{fig16} in
order of their Right Ascension.

Some of the results of our spectroscopic observations are summarized
in the color-color diagram shown in Figure \ref{fig10}, where we
identify the various spectral types of the 76 confirmed white dwarfs
with available SDSS colors.  As expected, the DQ stars are located in
the appropriate region of the ($u-g$, $g-r$) diagram, and in the next
phase of the survey, we plan to use this characteristic to identify
all possible DQ stars in SUPERBLINK. We also note the presence of a DA
+ dM system at $(u-g,g-r)\sim(2.8,1.2)$, in the redder part of the
diagram. Finally, the sample of DC stars follows the theoretical, pure hydrogen 
sequence, with only one outlier near $u-g=2.0$, giving
us a preliminary indication of the atmospheric composition even before
performing a full analysis of their energy distribution.

In the following section, we present a preliminary spectroscopic
analysis of the DA component of our survey.

\section{Atmospheric Parameter Determination of DA Stars}

The coolest white dwarfs in our sample are either featureless, or
present too few spectral lines for a proper spectroscopic analysis,
and the determination of their atmospheric parameters ($\Te$, $\logg$)
can only be achieved from an analysis of their photometric energy
distribution (see, e.g., \citealt{BRL}). At the moment, not enough
photometric information is available to proceed with a homogeneous
analysis of the coolest objects in our sample, and we are still
securing the appropriate optical and infrared photometry for cool DA,
DC, DQ, and DZ stars, the results of which will be reported in subsequent papers. 
We thus restrict our determination of the atmospheric parameters
to the subsample of 84 spectroscopically confirmed DA stars for
which the spectroscopic technique can be successfully applied.

\subsection{Theoretical Framework}

Our model atmospheres and synthetic spectra for DA stars are built
from the model atmosphere code originally described in \citet{BSW95}
and references therein, with recent improvements discussed in
\citet{tb09}. These are pure hydrogen, plane-parallel model
atmospheres, with non-local thermodynamic equilibrium effects
explicitly taken into account above $\Te=30,000$~K, and energy
transport by convection is included in cooler models following the
ML2/$\alpha=0.7$ prescription of the mixing-length theory. The
theoretical spectra are calculated within the occupation formalism of
\citet{HM88}, which provides a detailed treatment of the level
populations as well as a consistent description of bound-bound and
bound-free opacities. We also rely on the improved calculations for
the Stark broadening of hydrogen lines from \citet{tb09}, which include
nonideal perturbations from protons and electrons directly inside the
line profile calculations.  Our model grid covers a range of effective
temperature between $\Te = 1500$ K and 120,000 K, and $\logg$ values
between 6.0 and 9.5.

Our fitting technique is based on the approach pioneered by
\citet[][see also \citealt{liebert05}]{bergeron92}, which relies on
the nonlinear least-squares method of Levenberg-Marquardt
\citep{press86}. The optical spectrum of each star, as well as the
model spectra (convolved with a Gaussian instrumental profile), are
first normalized to a continuum set to unity. The calculation of $\chi
^2$ is then carried out in terms of these normalized line profiles
only. The atmospheric parameters -- $\Te$, $\logg$ -- are considered
free parameters in the fitting procedure.

Special care needs to be taken in the case of DA stars with an
unresolved M dwarf companion in order to reduce the contamination of
the white dwarf spectrum by the companion.  When the contamination
affects only \hb, and sometimes $\rm{H}{\gamma}$ as well, we simply
exclude these lines from the fit (e.g., 05280+4820 and 23229+3358). At
other times, emission lines from the M dwarf are also observed in the
center of the Balmer lines, in which case the line centers are also simply 
excluded from our fitting procedure (e.g., 04586+6209). A similar
approach was also adopted if the flux contribution from the M dwarf is
too important and ``fills up'' the Balmer line cores, resulting in
predicted lines that are too deep (06556+5920, 16171+0530, and
23283+3319). In some cases,
however, the white dwarf spectrum is too contaminated by the M dwarf
companion to be fitted with the simple approach described above (e.g.,
04032+2325E --- not to be confused with the DA star 04032+2325W --- and
11036+1555\footnote{In the case of 11036+1555, we even detect the 4226
  \AA\ line from the M dwarf in the white dwarf spectrum.}), and a
more robust fitting procedure using M dwarf templates will be required
\citep{gianninas2011}.  These results will be presented elsewhere.

\subsection{Spectroscopic Results}

Even though the spectroscopic technique is arguably the most accurate
method for measuring the atmospheric parameters of DA stars, it has an
important drawback at low effective temperatures ($T_{\rm eff}\lesssim
13,000$~K) where spectroscopic values of $\log g$ are significantly
larger than those of hotter DA stars, the so-called high-$\log g$
problem (see \citealt{tremblay2010} and references therein).
\cite{tremblay3D} showed that this high-$\logg$ problem is actually
related to the limitations of the mixing-length theory used to
describe the convective energy transport in DA stars, and that more
realistic, 3D hydrodynamical model atmospheres are required in order
to obtain a surface gravity distribution that resembles that of hotter
radiative-atmosphere DA stars. Since these spurious high-$\log g$
values affect directly the estimated distances, \citet{noemi2012}
derived an empirical procedure (see their Section 5 and Figure 16) to
correct the $\log g$ values based on the DA stars in the Data Release
4 of the Sloan Digital Sky Survey, analyzed by
\citet{tremblay2011}. We adopt a similar approach here and apply their
$\logg$ correction to all DA stars between $\Te=7000$~K and 14,000~K.

\begin{figure*}[t!]
\epsscale{0.8}
\plotone{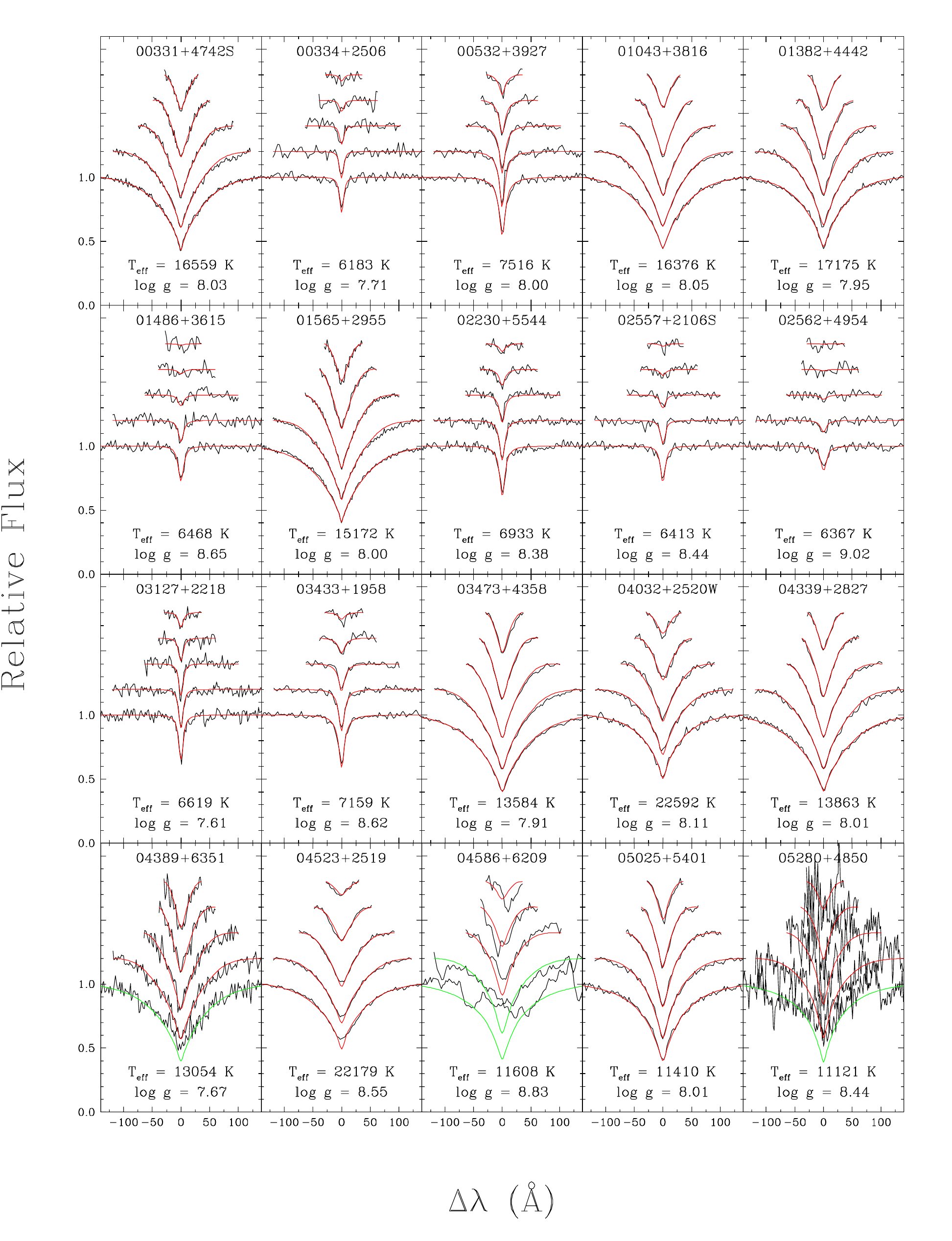}
\caption{(a) - Fits to the optical spectra of the DA stars in
  our sample. The lines range from $\rm{H}\beta$ (bottom) to
  {\rm{H}8} (top), each offset vertically by a factor of
  0.2. Theoretical line profiles shown in green are not used in the
  fitting procedure. \label{fits}}
\end{figure*}

\addtocounter{figure}{-1}
\begin{figure*}[t!]
\epsscale{0.8}
\plotone{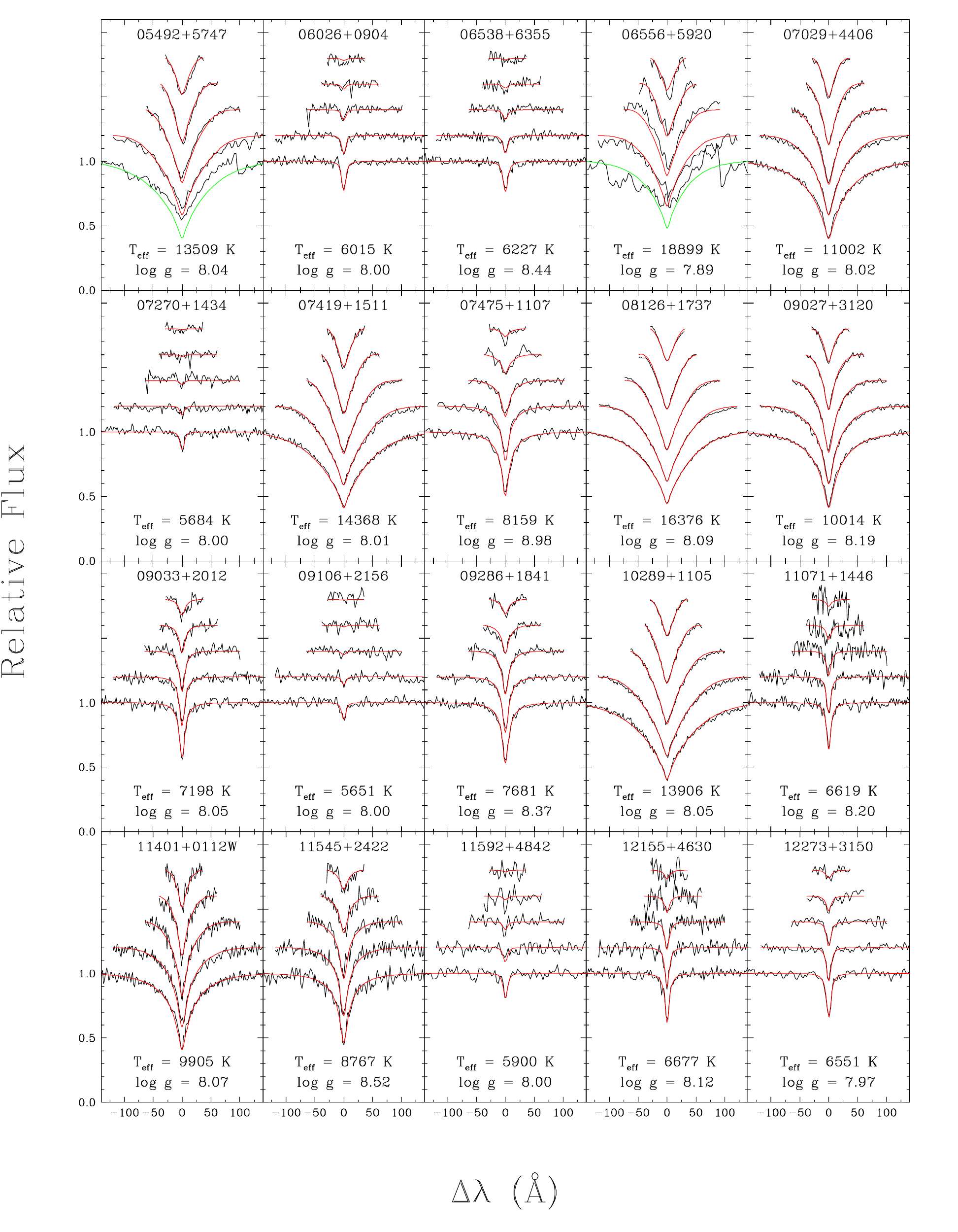}
\caption{(b) -continued.}
\end{figure*}

\addtocounter{figure}{-1}
\begin{figure*}[t!]
\epsscale{0.8}
\plotone{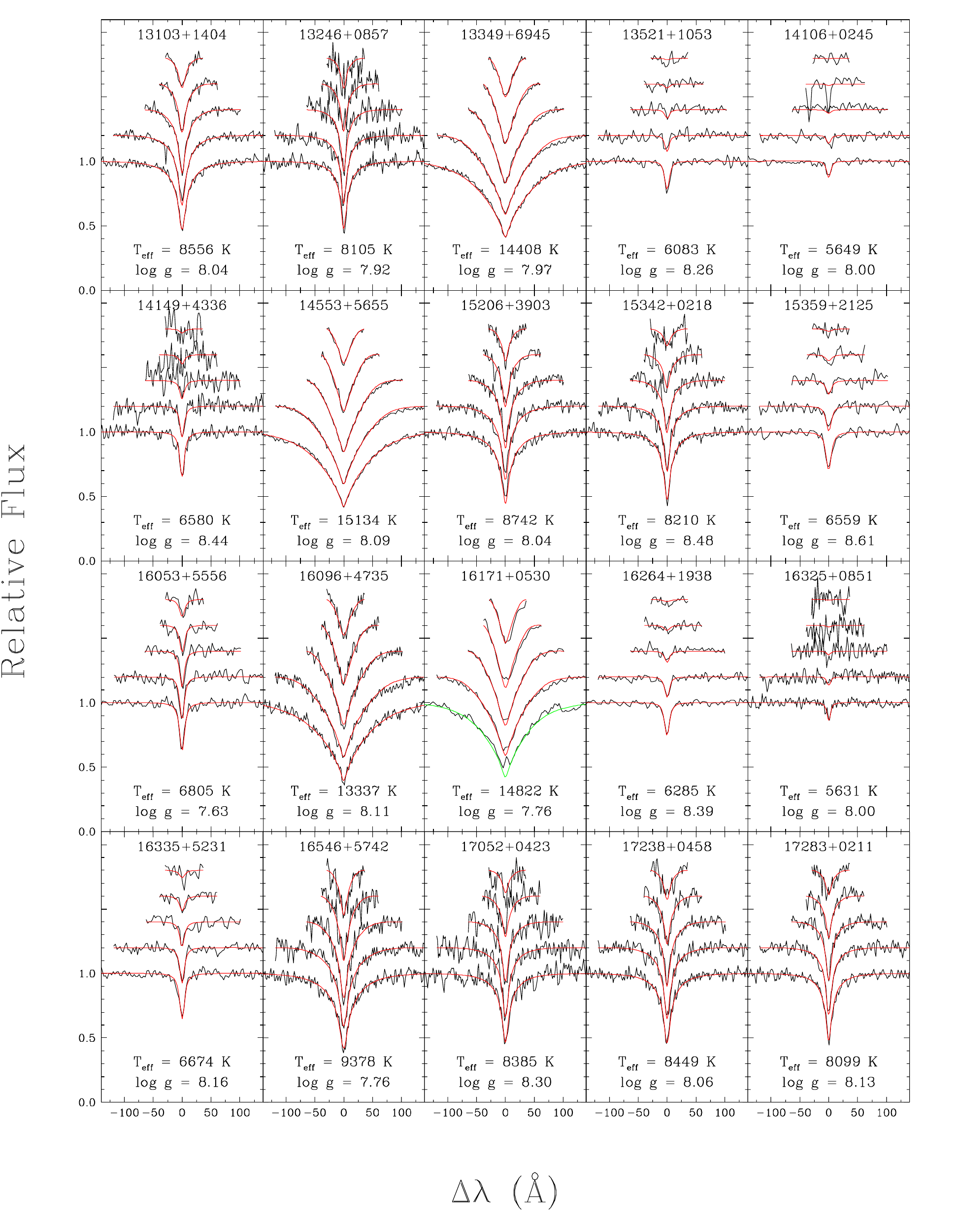}
\caption{(c) - continued}
\end{figure*}

\addtocounter{figure}{-1}
\begin{figure*}[t!]
\epsscale{0.8}
\plotone{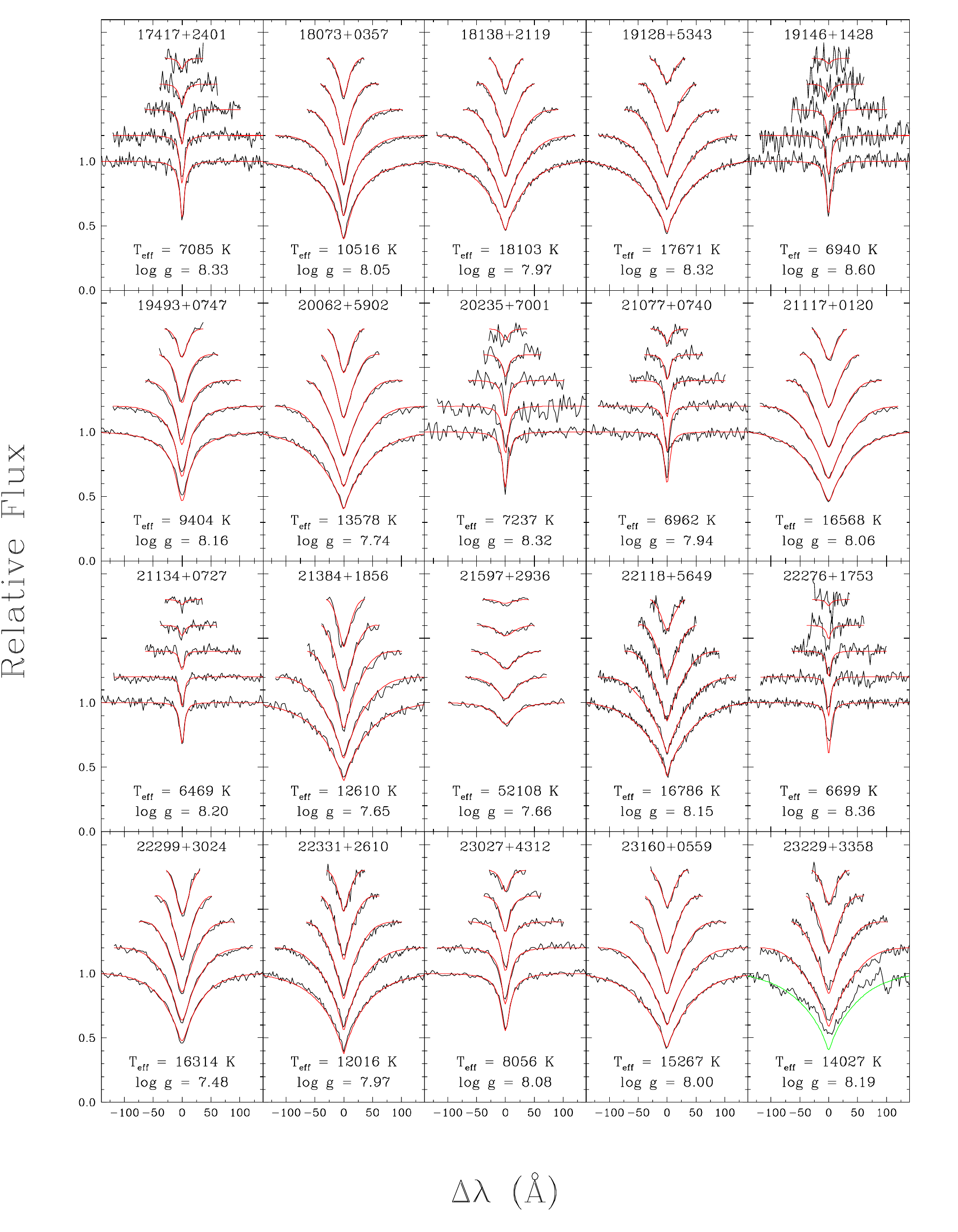}
\caption{(d) - continued}
\end{figure*}

\addtocounter{figure}{-1}
\begin{figure*}[t!]
\epsscale{0.8}
\plotone{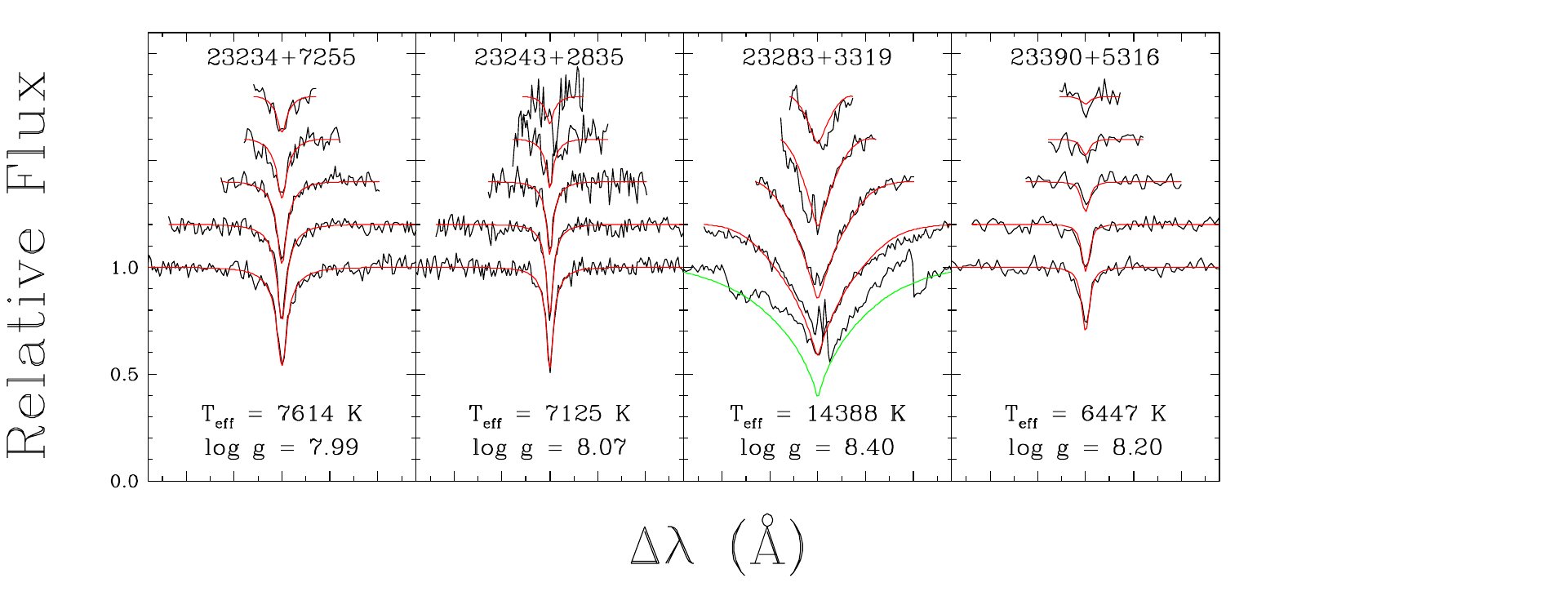}
\caption{(e) - continued}
\end{figure*}

The spectroscopic fits for our subsample of 84 DA stars
are displayed in Figure \ref{fits}. The corresponding atmospheric
parameters ($\Te$ and $\logg$) are reported in Table \ref{tab5}
together with the stellar mass ($M/$\msun), absolute absolute visual
magnitude ($M_V$), luminosity ($L/L_\odot$), estimated visual
magnitude ($V$), spectroscopic distance ($D$), and white dwarf cooling
time ($\tau$).  Whenever necessary, we rely on the same evolutionary
models as those described above to derive these quantities. In
principle, the spectroscopic distance can be obtained directly from
the distance modulus, by combining the theoretical absolute magnitude
in a single given bandpass with the observed magnitude in the same
bandpass. However, since the photometric errors can be large in some
systems we used --- the USNO photographic magnitudes in particular
---, we estimated the spectroscopic distances by using the full set of
photometry available for each star, and calculated an average
spectroscopic distance, properly weighted by the photometric
uncertainties in each bandpass. This is equivalent to using the
photometric method described above but by forcing the effective
temperature at the spectroscopic value, thus fitting only the solid
angle $\pi(R/D)^2$, where $R$ is the radius of the star determined
from the spectroscopic $\log g$ value. In doing so, we also fold in
the uncertainty of the spectroscopic $\log g$ measurement.

\begin{figure}[h!]
\epsscale{1.3}
\plotone{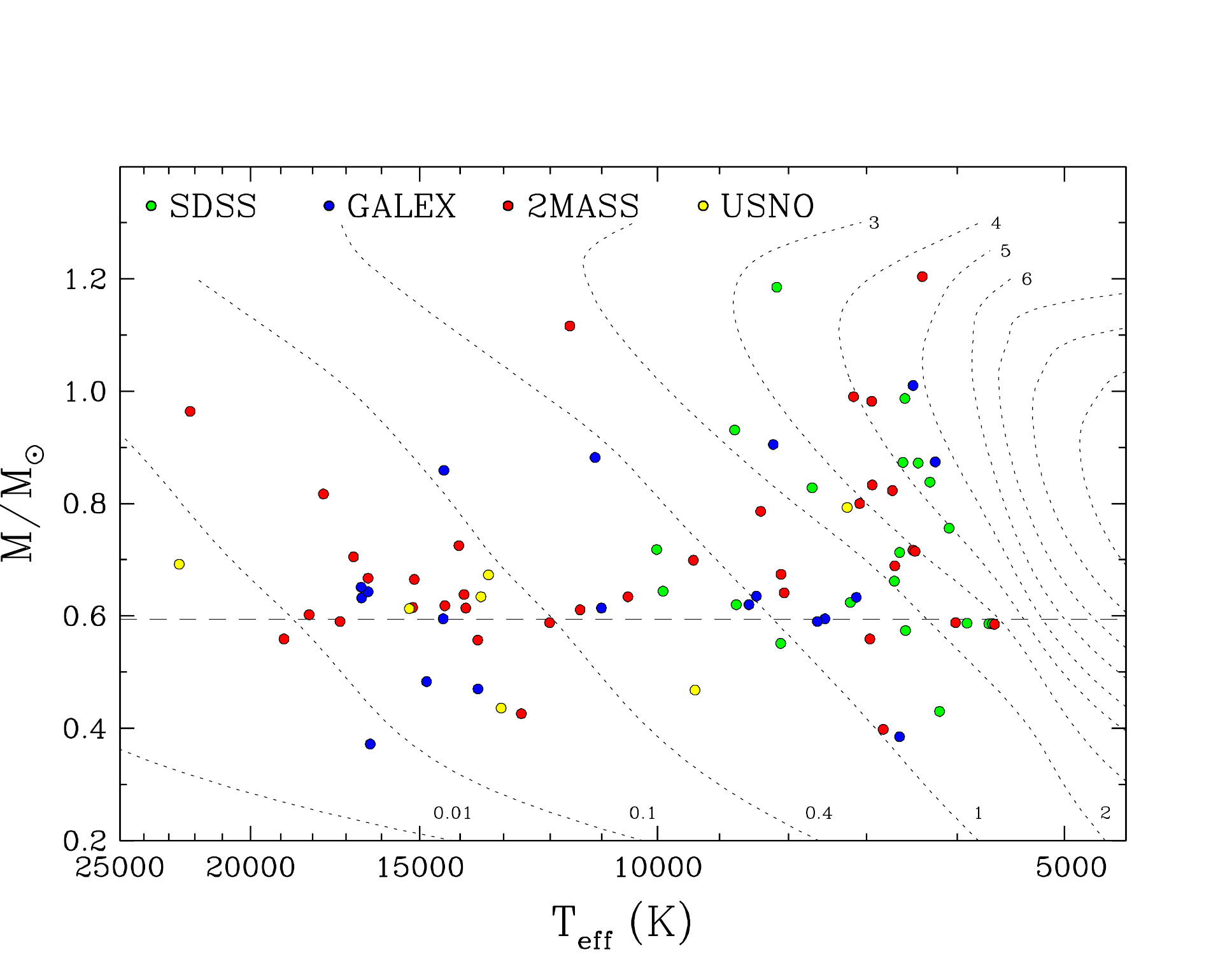}
\caption{Mass as a function of effective
  temperature for a subsample of 84 DA white dwarfs with spectroscopic
  mass determinations. All stars are identified with a different color
  based on the photometric system from which they were discovered.
  The dotted lines represent the theoretical isochrones for our C/O
  core evolutionary models with thick hydrogen layers, corresponding
  to the white dwarf cooling age in units of Gyr. The dashed line
  indicates the median mass of DA white dwarfs, 0.594 $M_{\odot}$, as
  determined by \citet{tremblay2011}. \label{fig18}}
\end{figure}

The mass distribution for the DA stars in our sample is displayed in
Figure \ref{fig18} as a function of effective temperature.  This
figure clearly illustrates the efficiency of our survey to identify
white dwarfs using reduced proper motion diagrams even at very low
effective temperatures. We also distinguish with various color codes
the criteria used in our survey to discover each white dwarf, allowing
us to study the impact of one particular photometric system on the
selection process as a function of temperature. For instance, white
dwarfs identified on the basis of GALEX photometry extend down to
relatively low effective temperatures.  Indeed, the observed
photometric sequence allows us to apply our selection criteria down to
${\rm NUV}-V=6.5$ (see Figure \ref{fig6}), or $\Te\sim
5300$~K. Similarly, white dwarfs identified on the basis of $ugriz$
photometry are mostly found at the low end of the temperature
distribution. Most SDSS targets are intrinsically faint, and thus
include an impressive amount of cool white dwarfs that can only only
be identified through the use of reduced proper motion
diagrams. Surprisingly, white dwarfs identified on the basis of 2MASS
photometry are found at all temperatures. This is due to the fact that
our photometric sequences allow us to apply our color criteria as blue
as $V-J=-0.5$ (see Figure \ref{fig7}), or $\Te\sim20,000$~K.
Finally, only a few white dwarfs in this subsample were identified on
the basis of USNO photographic magnitudes. From these results, we can
conclude that even though SDSS represents the most reliable
photometric data set, GALEX, 2MASS, and even photometric magnitudes
are also required to identify white dwarfs over the complete range of
effective temperature.

\begin{figure}[h!]
\epsscale{1.3}
\plotone{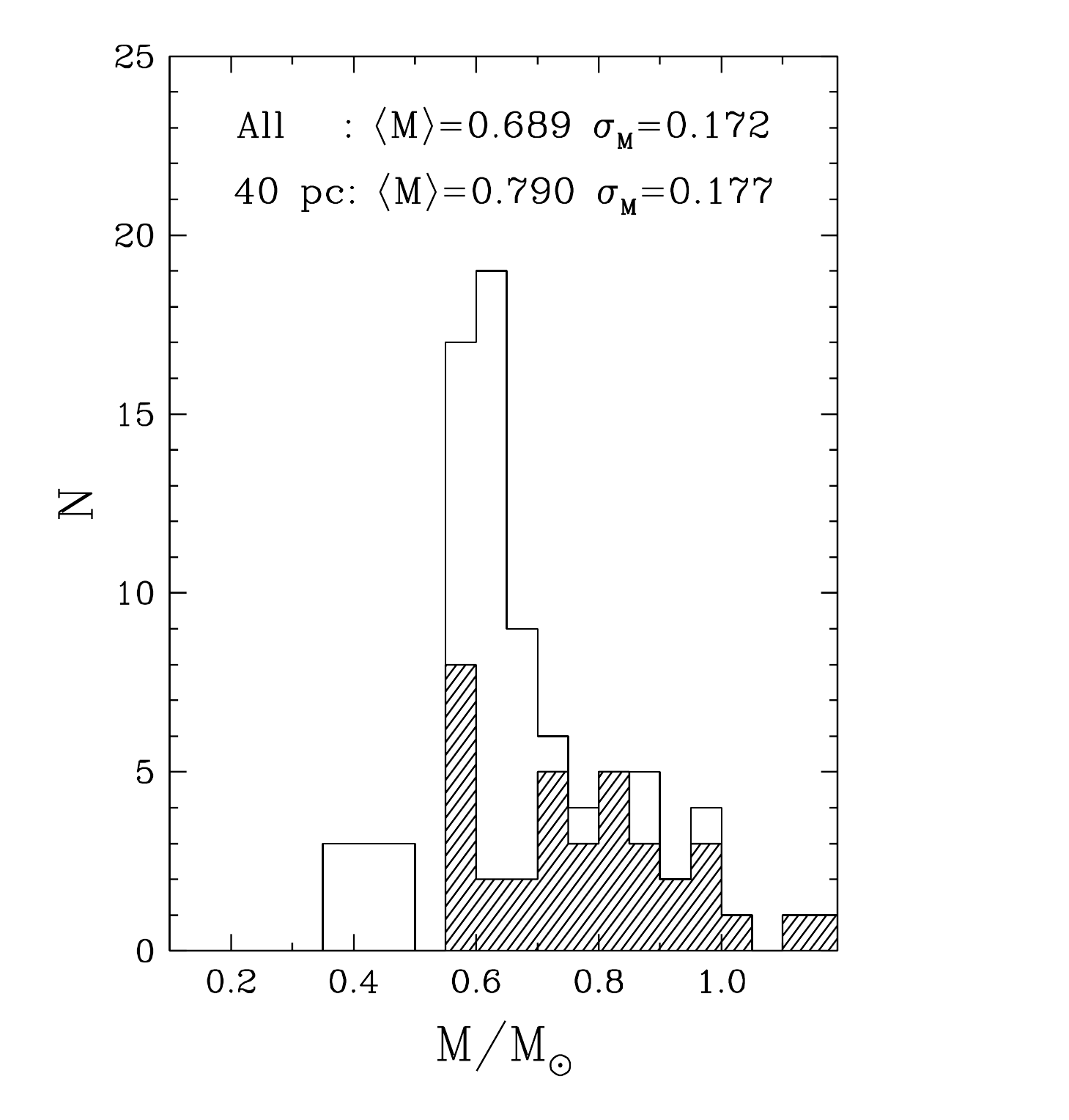}
\caption{Mass distribution for the DA white dwarfs
  in our sample. The solid line histogram shows the
  distribution for the 84 DA stars with spectroscopic masses available,
  while the shaded histogram corresponds to the subsample of 37 objects
  with spectroscopic distances less than 40 pc from the Sun.
  The mean values and standard deviations are given in the figure.
  \label{fig19}}
\end{figure}

The mass distribution of DA white dwarfs in our subsample, regardless
of their temperature, is displayed in Figure \ref{fig19}.  The
mean mass of these 84 DA stars is 0.689 \msun\ with a standard
deviation of $\sigma=0.172$ \msun, a value significantly larger than
the value obtained by \citet{noemi2012} for the DA white dwarfs within
20 pc of the Sun (0.647 \msun\, with $\sigma=0.171$ \msun). One obvious
difference is that we do not include here the white dwarfs already
known in the literature. Most likely these are brighter, intrinsically
more luminous, and probably have larger radii and thus lower
masses. The mass distribution of the 37 DA white dwarfs within 40 pc
of the Sun displayed in Figure \ref{fig19} (shaded histogram)
actually shows an important high-mass component (see also Figure
\ref{fig18}).  These high-mass white dwarfs, with their small
stellar radii, are intrinsically less luminous than their normal-mass
counterparts, and they are thus more abundant in a volume-limited
sample, such as the local neighborhood, than in a magnitude-limited
sample. Our results indicate that we are successfully recovering these
high-mass white dwarfs in our survey, often missing in
magnitude-limited surveys (see, e.g., \citealt{liebert05} in the case
of the PG survey).

\section{Discussion}

\subsection{Comparison of Spectroscopic and Photometric Distances}

During the target selection process, distances were
estimated using approximate $V$ magnitudes together with various
color-magnitude relations, displayed in Figures \ref{fig5} to
\ref{fig8}. These distance estimates were later improved by
comparing theoretical average fluxes to the set of available
photometry, properly weighted by their uncertainty. At that point, we
simply assumed a surface gravity of $\log g=8.0$, and considered both
$\Te$ and the solid angle as free parameters.  These estimates are
referred to as {\it photometric distances}. The spectroscopic
analysis, on the other hand, provides {\it spectroscopic distances},
where for a given star, theoretical absolute magnitudes are computed
from the spectroscopic values of $\Te$ and $\log g$, and compared to
the set of available photometry. In both cases, if only the $V$
photographic magnitude is available, the 0.5 magnitude error will
introduce a 23\% uncertainty on the estimated spectroscopic
distance. If additional photometry is available, however, this
distance uncertainty can be significantly reduced (see Table
\ref{tab5}).

\begin{figure}[h!]
\epsscale{1.6}
\plotone{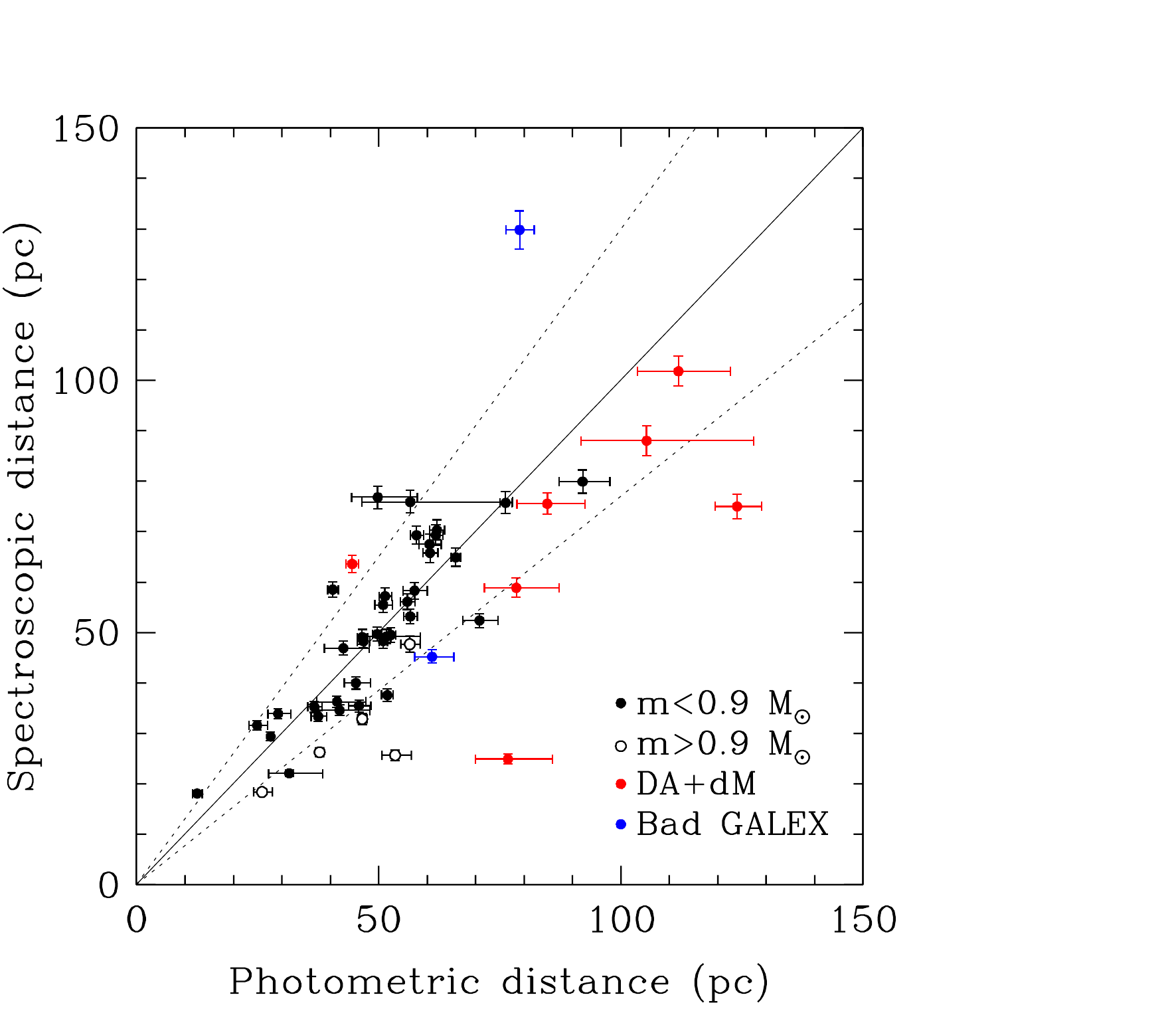}
\caption{Comparison of photometric and
  spectroscopic distances for white dwarfs with $\Te>7000$~K, as
  defined in the text. The solid line represents the 1:1
  correspondence, while the dotted lines represent a $\pm23\%$
  dispersion. \label{fig20}}
\end{figure}

The comparison between photometric and spectroscopic distances for the
DA white dwarfs in our sample is displayed in Figure \ref{fig20}. We
restrict this comparison to $\Te>7000$~K since the Balmer lines in
cooler objects become too weak to be fitted properly with the
spectroscopic method, yielding spurious $\log g$ values at low
temperatures (see Figure \ref{fig18}) and corresponding distances. The
dotted lines in Figure \ref{fig20} represent a $\pm23\%$ difference
between both estimates (i.e., the maximum error on spectroscopic
distances obtained from photographic magnitudes, as discussed in the
previous paragraph). The bulk of stars is generally found within these
limits. Part of the observed dispersion in Figure \ref{fig20} can be
attributed to the intrinsic mass distribution of our sample. Indeed,
all our color-magnitude calibrations assumed a typical mass of 0.6
\msun, but as shown in Figures \ref{fig5} to \ref{fig8}, there is an
intrinsic dispersion in absolute magnitude due to the mass (or radius)
distribution of white dwarfs. In particular, white dwarfs with very
high ($M>0.9$ \msun) spectroscopic masses yield photometric distances
that are overestimated; these are identified with a different symbol
in Figure \ref{fig20}.

Another source of scatter is due to the presence of M dwarf
companions, which make the system brighter at visual and infrared 
magnitudes compared with single DA stars. 
Since these magnitudes were used to estimate their photometric
distance ($M_V$ vs $V-J$), this can easily account for the large
discrepancies with spectroscopic distances. Indeed, in the
spectroscopic distance calculation, the less accurate magnitudes weigh
less, and the more accurate $JHK$ photometry dominates the
distance solution. Finally, as noted in Table \ref{tab5}, we have certain
doubts about the cross-correlation with the GALEX database for a handful of
stars in our sample. For these objects, the GALEX photometry is
inconsistent with the rest of the spectral energy distribution, and they had to
be omitted from the fits used to estimate the spectroscopic
distances. However, as in the previous cases, these colors were used
to obtain our initial distance estimate.

To summarize, most objects in Figure \ref{fig20} are found between
the $\pm23\%$ dispersion in distance, and the stars falling outside
these limits can be separated into three categories: DA stars with M
dwarf companions, high-mass white dwarfs, and stars with large
photometric uncertainties (see the corresponding error bars in Figure
\ref{fig20}). The previously estimated 15 pc error is thus
enough to identify white dwarfs with reasonably accurate photometry, 
and we thus conclude that searching at 55 pc in order
to find all white dwarfs within 40 pc is realistic, especially when
photometry such as SDSS, GALEX, or 2MASS is available.

Our preliminary spectroscopic analysis of DA stars presented in Table
\ref{tab5} yields 11 white dwarf candidates within 25 pc of the
Sun, including 5 candidates within the 20 pc sample.  Incidentally, a
few of these objects already have a parallax measurement
available. Indeed, 21134+0727 (G25-20) has a parallax from
\citet{dahn88}, $\pi=0\farcs0411 \pm 0\farcs0038$ yr$^{-1}$, placing
it within 25 pc. Also, if 22118+5649 is a common proper motion
companion to LTT 16500, as it is suspected to be (Subasavage et
al.~2012, private communication), then it has a parallax of
$\pi=0\farcs02677 \pm 0\farcs00018$ yr$^{-1}$ (or $D=37.4$ pc), and
thus not a member of the 20 pc sample, while still within 40 pc of the
Sun. Finally, a private communication from J.~Subasavage confirms that
16325+0851 is indeed within 25 pc of the Sun. So even though
spectroscopic distances are more accurate than the previous
photometric estimates, the only way to confirm the membership of white
dwarfs to the local sample is through trigonometric parallax
measurements. Such measurements would not only provide reliable
distances, but would also yield mass determinations for the coolest
objects in our sample analyzed with the photometric technique.

\subsection{Success Rate of Discovery}

\begin{figure}[t!]
\epsscale{1.2}
\plotone{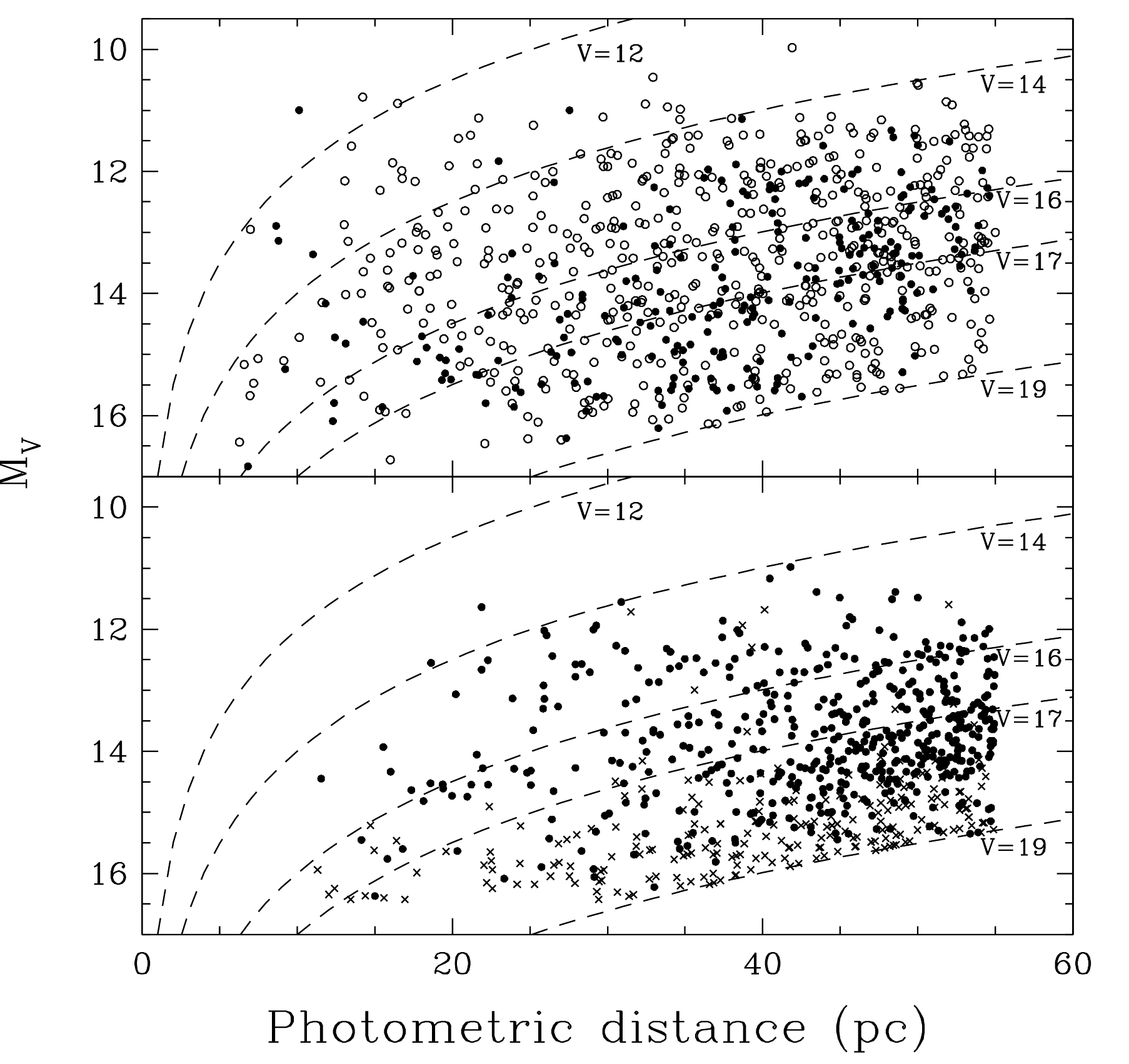}
\caption{Absolute magnitude as a function of
  photometric distance. In the upper panel, the filled circles
  represent the 193 new white dwarfs identified in our survey, while
  the open circles correspond to the 499 white dwarfs already known in
  the literature. Also shown by dashed lines in the figure are lines
  of constant apparent $V$ magnitudes. The white dwarf candidates in
  our survey without spectroscopic confirmation are shown in the lower
  panel. The lower-priority candidates (those identified on the basis
  USNO photographic magnitudes) are shown with cross
  symbols. \label{fig21}}
\end{figure}

The absolute visual magnitudes
(estimated from the calculated $V$ magnitudes and photometric
distances) for the 193 new white dwarfs identified in our survey are
plotted in the upper panel of Figure \ref{fig21} as a function of
photometric distance. Also shown are the 499 white dwarfs in
SUPERBLINK already known in the literature. The candidates still
without spectroscopic confirmation are displayed separately in the
lower panel; the objects selected on the basis of their USNO
photographic magnitudes are considered second priority targets because
of their higher probability of being contaminants from the main
sequence. In each panel, the dashed lines represent lines of constant
apparent $V$ magnitude. We note that the white dwarfs identified in
our survey are dominated by objects fainter than $V=16$, and that most
of them are found at photometric distances larger than 20 pc. This is
not surprising since the census of white dwarfs within 20 pc of the
Sun is believed to be at least $90\%$ complete
\citep{noemi2012}. There are still a few white dwarf candidates on our
target list within 20 pc that have no spectroscopic data, due to
observational constraints, but these stars are currently on our high
priority list.

From the results shown in the upper panel of Figure \ref{fig21}, we
can determine that the ratio of new to known white dwarfs is
$193/499\sim39\%$. Also, out of the 286 candidates observed, 220 are
confirmed white dwarfs (27 in the literature\footnote{Some spectra of
  spectroscopically confirmed white dwarfs were secured by us and will
  be used in our next paper as part of a study of the total white
  dwarf content of SUPERBLINK within 40 pc of the Sun.} and 193 in our
survey), for a success rate of $77\%$.  This number is close to the
$80\%$ expected from our selection criteria, and we conclude that our
survey is quite efficient for recovering the missing fraction of white
dwarfs in the solar neighborhood.

The lower panel of Figure \ref{fig21} reveals that a significant
fraction of our remaining white dwarf candidates are fainter than $V=17$ (590
objects fainter versus 329 objects brighter). The spectroscopic
identification of these stars with 2 to 4-m telescopes requires
integration times on the order of an hour under excellent weather
conditions. The candidates deserving spectroscopic follow-up must then
be carefully chosen, and our high-priority list now includes 89 of
these faint candidates and 186 ``bright'' targets, for a total of 275
high-priority targets, excluding 120 objects we already observed after
2010 October and that are still being reduced. Future observations
 will be dedicated to the follow-up of these high-priority
white dwarf candidates, in particular those identified on the basis of
SDSS or GALEX photometry.

\subsection{Increasing the completeness of the current census}
\begin{figure}[h!]
\epsscale{1.2}
\plotone{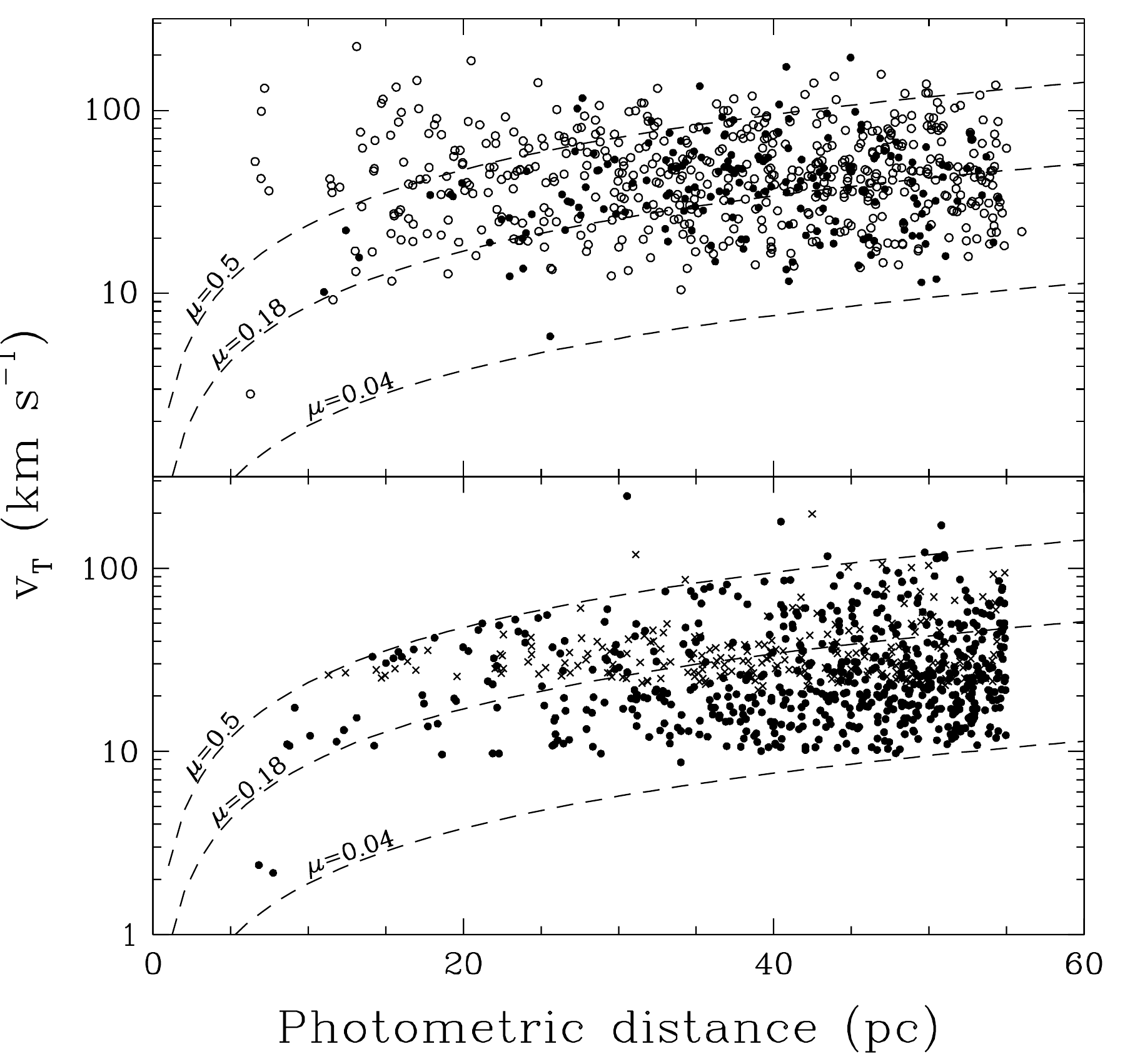}
\caption{Transverse velocity as a function of
  distance, showing the kinematic bias due to the proper motion limit
  of the SUPERBLINK catalog ($\mu > 40$ mas $\rm{yr}^{-1}$).  In the
  upper panel, the filled circles represent the 193 new white dwarfs
  identified in our survey, while the open circles correspond to the
  499 white dwarfs already known in the literature.  Also shown by
  dashed lines in the figure are lines of constant apparent proper
  motion (in units of arcsec yr$^{-1}$).  The white dwarf candidates
  in our survey without spectroscopic confirmation are shown in the
  lower panel. The lower-priority candidates (those identified on the
  basis USNO photographic magnitudes) are shown with cross
  symbols.\label{fig22}}
\end{figure}

We have already established the success rate of our spectroscopic
survey, and we are now interested in its completeness. First of all,
our white dwarf sample is directly affected by the completeness of the
SUPERBLINK catalog, which is high because of its low proper motion
limit ($\mu > 0\farcs04$ yr$^{-1}$) which minimizes the kinematics
bias. To illustrate the effect of proper motion on kinematics, we plot
in Figure \ref{fig22} the transverse motions $v_t$ (i.e.~the projected
motions on the plane of the sky, where $v_t=4.47{\mu}d$) for all stars
in our sample as a function of the photometric distance $D$, as
calculated in Section 4.1. As explained in \citet{LG2011}, a star at
50 pc from the Sun with $\mu > 0\farcs04$ yr$^{-1}$ has a transverse
velocity $v_t < 9.48$ km s$^{-1}$, which will occur with a probability
of about $10\%$ for stars in the solar neighborhood (see their Section
2.2 and Figure 1). Their diagram shows that operating with a proper
motion limit of $\mu > 0\farcs150$ yr$^{-1}$ will only detect half of
the stars at 40 pc and very few stars (only those with very large
components of motion) at 100 pc. However, a sample with a proper
motion limit $\mu > 0\farcs04$ yr$^{-1}$ will include $\sim95\%$ of
the stars at 40 pc and $\sim70\%$ of the stars at 100 pc.

Hence, in terms of new white dwarf identification as a function of
proper motion, we find that for $\mu> 0\farcs5$ yr$^{-1}$ (see
corresponding dashed line in Figure \ref{fig22}), which corresponds to
the limit of the LHS survey, the ratio of new to known white dwarfs is
$8.8\%$, while this ratio reaches 43.5\% for $0\farcs5>\mu>0\farcs18$
yr$^{-1}$ (where the lower proper motion limit is that of the NLTT
survey), and it then drops slightly to $41.3\%$ for
$0\farcs18<\mu<0\farcs04$ yr$^{-1}$.  Our survey is thus more
efficient for proper motions lower than the LHS limit, but our results
also demonstrate that the sample of white dwarfs with $\mu>0\farcs5$
yr$^{-1}$ could host up to 7 more white dwarfs. Previous searches for
white dwarfs within the NLTT limit were also incomplete, since 33 of
our new identifications have an NLTT designation. Note also that the
NLTT and LHS appear to be $\gtrsim80\%$ complete down to the 19th
magnitude, but only in the Completed Palomar Region (CPR), i.e.~for
$\delta>-32.5^\circ$ and outside a band $\pm10^\circ$ of the Galactic
plane \citep{lspm05}.

Spectroscopic distances were obtained for 84 out of the 193 newly
identified white dwarfs, while a preliminary photometric analysis (not
presented here) of a subsample of the coolest objects was performed
for another 78 white dwarfs. From this combined analysis of 162 white
dwarfs, we find that 126 objects are within 55 pc of the Sun, and 93
within 40 pc. The spectroscopic analysis of 1151 DA stars by
\citet{gianninas2011} contains 223 white dwarfs from the WD Catalog
located in the northern hemisphere whose spectroscopic distances are
within 55 pc from the Sun, and 121 within 40 pc. Using this latter
survey, our ratio of new to known white dwarfs is estimated at $56\%$
within 55 pc, and $77\%$ within 40 pc. This difference in ratio will
most likely be reduced when all candidates between 40 and 55 pc are
observed. It was also
mentioned in Section 7.2 that the ratio of new to known white dwarfs
from the literature was $39\%$, while our success rate in detecting
white dwarfs (both new and known) is $77\%$. Our survey is thus
efficient for recovering white dwarfs that are already found in the
literature as well as new identifications.

In spite of the success of our survey, the first sample of newly
identified white dwarfs presented in this paper is far from complete,
but the survey has not reached its limit yet. Our analysis represents
the first results of an ongoing effort, and more data are still being
collected and analyzed. Moreover, SUPERBLINK is currently being
cross-correlated with the SDSS DR7 and GALEX GR7, providing additional
high-quality photometric information to replace USNO magnitudes in our
selection process. This will eventually result in more high-priority
candidates, and will also help in the identification of DQ and DZ
stars, which separate well in color-color diagrams, as we showed
earlier, but only when such color information is available. Finally,
these new magnitudes may also complete the set of photometry for white
dwarfs in the SUPERBLINK catalog, allowing fits to the energy
distribution of the cool white dwarfs that cannot be analyzed
spectroscopically.  Our future catalog will provide more candidates
for parallax measurements, as well as more cool, massive, magnetic,
and astrophysically challenging white dwarfs, while being at least
$80\%$ complete within 40 pc of the Sun.  We will also be able to
provide statistics on the Solar Neighborhood based on a sample of
white dwarfs large enough to reduce the uncertainties related to small
number statistics.
 
\acknowledgements We would like to thank the director and staff of
Steward Observatory and Kitt Peak National Observatory for the use of
their facilities.  This work was supported in part by the NSERC Canada
and by the Fund FRQ-NT (Qu\'ebec). S.L. was supported in this research
by NSF grants AST-0607757 and AST-0908406. S.L also acknowledges
support for this work from the GALEX Guest Investigator program under
NASA grant NNX09AF88G. This research has made use of the SIMBAD
database, and the VizieR catalog access tool, operated at CDS,
Strasbourg, France. This publication also makes use of data products
from the Two Micron All Sky Survey, which is a joint project of the
University of Massachusetts and the Infrared Processing and Analysis
Center/California Institute of Technology, funded by the National
Aeronautics and Space Administration and the National Science
Foundation.


\bibliographystyle{apj} 

\clearpage

\clearpage
\LongTables 
\begin{deluxetable}{lcccccrrc}
\tabletypesize{\scriptsize}
\tablecolumns{9}
\tablewidth{0pt}
\tablecaption{Spectroscopically Confirmed White Dwarfs from SUPERBLINK $-$ Astrometry}
\tablehead{
\colhead{ } &
\colhead{ } &
\colhead{ } &
\colhead{RA} &
\colhead{DEC} &
\colhead{$\mu_{\rm{tot}}$} &
\colhead{$\mu$RA} &
\colhead{$\mu$DE} &
\colhead{ }\\
\colhead{PM I} &
\colhead{NLTT} &
\colhead{SDSS} &
\colhead{($J_{2000}$)} &
\colhead{($J_{2000}$)} &
\colhead{$(\arcsec \rm{yr}^{-1})$} &
\colhead{$(\arcsec \rm{yr}^{-1})$} &
\colhead{$(\arcsec \rm{yr}^{-1})$} &
\colhead{Notes}}
\startdata
00023$+$6357 &      &                &00:02:22.61 &  +63:57:44.3& 0.911  & 0.906  & 0.092  &\\
00079$+$3947 &  317 &                &00:07:54.11 &  +39:47:32.2& 0.370  & 0.362  &$-$0.075&\\
00217$+$2640 & 1119 &                &00:21:47.32 &  +26:40:36.3& 0.330  &$-$0.074&$-$0.322&\\
00276$+$0542 &      &                &00:27:36.63 &  +05:42:03.2& 0.381  & 0.266  &$-$0.273&\\
00331$+$4742S&      &                &00:33:10.51 &  +47:42:12.4& 0.079  &$-$0.060& 0.052  &\\
\\
00334$+$2506 & 1783 &J003325.44+25061&00:33:25.28 &  +25:06:14.1& 0.499  & 0.477  & 0.146  &\\
00532$+$3927 &      &                &00:53:13.75 &  +39:27:06.1& 0.209  &$-$0.152&$-$0.143&\\
\enddata
\tablecomments{
(1) Also in Vennes et al.~(2011);  (2) Classified using our own data, but SDSS spectrum exists; (3) Also in Kilic et al.~(2010); (4) Also in Sayres et al.~(2012); (5) Also in Tonry et al.~(2012).}
\label{tab3}
\end{deluxetable}

\clearpage
\begin{landscape}
\begin{deluxetable}{lcccccccccccccclc}
\tabletypesize{\scriptsize}
\tablecolumns{20}
\tablewidth{0pt}
\tablecaption{Spectroscopically Confirmed White Dwarfs from SUPERBLINK $-$ Photometry}
\tablehead{
\colhead{PM I} &
\colhead{FUV} &
\colhead{NUV} &
\colhead{$B_F$} &
\colhead{$R_I$} &
\colhead{$I_N$} &
\colhead{$J$} &
\colhead{$H$} &
\colhead{$K_S$} &
\colhead{$u$} &
\colhead{$g$} &
\colhead{$r$} &
\colhead{$i$} &
\colhead{$z$} &
\colhead{$\rm{D_{phot}}$ (pc)} &
\colhead{ST} &
\colhead{Notes}}
\startdata
    00023$+$6357 &$-$  &$-$  &$-$ &16.2&17.3&15.80&15.57&15.51&$-$  &$-$  &$-$  &$-$  &$-$  & ~25.9&DC   &\\
    00079$+$3947 &$-$  &$-$  &17.4&16.2&15.8&15.18&14.85&14.65&$-$  &$-$  &$-$  &$-$  &$-$  & ~20.2&DC   &\\
    00217$+$2640 &$-$  &$-$  &17.9&17.2&16.9&16.18&15.91&15.81&$-$  &$-$  &$-$  &$-$  &$-$  & ~35.2&DC   &\\
    00276$+$0542 &$-$  &$-$  &17.2&15.6&15.4&14.97&14.67&14.57&$-$  &$-$  &$-$  &$-$  &$-$  & ~18.2&DA   &\\
    00331$+$4742S&14.55&14.63&14.9&14.9&$-$ &15.16&15.02&15.14&$-$  &$-$  &$-$  &$-$  &$-$  & ~50.8&DA   &\\
\\
    00334$+$2506 &$-$  &20.36&16.6&16.6&16.1&15.96&15.70&15.45&17.88&17.12&16.79&16.69&16.68& ~37.0&DA   &\\
    00532$+$3927 &$-$  &18.45&17.9&16.4&16.8&16.03&15.84&15.40&$-$  &$-$  &$-$  &$-$  &$-$  & ~54.4&DA   &\\
\enddata
\tablecomments{
(1) Classified using our own data, but SDSS spectrum exists; $~{\rm H} \alpha$ data will be obtained to confirm DC status (2)DA+DC candidate; (3) Magnetic candidate; (4) DA+DA candidate (5) Spectral range of $\lambda\lambda 3200-5300$; $~{\rm H}\alpha$ data will be obtained to confirm DC status. }\\
\label{tab4}
\end{deluxetable}
\clearpage
\end{landscape}

\clearpage
\begin{deluxetable}{lrlrcccrrc}
\tabletypesize{\scriptsize}
\tablecolumns{10}
\tablewidth{0pt}
\tablecaption{Atmospheric Parameters of DA White Dwarfs from SUPERBLINK}
\tablehead{
\colhead{PM I} &
\colhead{$\Te$ (K)} &
\colhead{log $g$} &
\colhead{$M/M_{\odot}$} &
\colhead{$M_V$} &
\colhead{log $L/L_{\odot}$} &
\colhead{$V^a$} &
\colhead{$D ({\rm pc})$} &
\colhead{log $\tau$} &
\colhead{Notes}}
\startdata
    00331+4742S & 16560 (125)     &8.03 (0.05)    &0.63 (0.03)    &11.14     &$-$1.96        &14.75       & 55.5 (1.5) &8.18           &          \\
    00334+2506  &  6180 (100)     &7.71 (0.27)    &0.43 (0.13)    &13.67     &$-$3.52        &16.43       & 44.6 (1.1) &9.15           &          \\
    00532+3927  &  7520 ( 44)     &8.00 (0.09)    &0.60 (0.05)    &13.26     &$-$3.33        &17.05       & 45.2 (1.3) &9.10           &    1     \\
    01043+3816  & 16380 (102)     &8.05 (0.05)    &0.64 (0.03)    &11.19     &$-$1.99        &15.35       & 58.3 (1.6) &8.21           &          \\
    01382+4442  & 17180 (157)     &7.95 (0.05)    &0.59 (0.03)    &10.96     &$-$1.85        &15.57       & 69.3 (2.0) &8.06           &          \\
\\
    01486+3615  &  6470 (138)     &8.65 (0.28)    &1.01 (0.17)    &14.87     &$-$4.01        &16.93       & 24.2 (0.9) &9.68           &          \\
    01565+2955  & 15170 (121)     &8.00 (0.05)    &0.62 (0.03)    &11.25     &$-$2.10        &14.64       & 48.3 (1.3) &8.28           &          \\
\enddata
\tablecomments{
(1) Bad GALEX photometry; (2) Binary but only one set of magnitudes; (3) Not in our new selection; (4) DA+dM. }
\tablenotetext{a}{Estimated from USNO photographic magnitudes.}
\label{tab5}
\end{deluxetable}

\end{document}